\documentclass[]{interact}

\usepackage{epstopdf}
\usepackage[caption=false]{subfig}

\usepackage[longnamesfirst,sort]{natbib}
\usepackage{graphicx}
\usepackage{multirow}
\usepackage{amssymb}
\bibpunct[, ]{(}{)}{;}{a}{,}{,}
\renewcommand\bibfont{\fontsize{10}{12}\selectfont}

\theoremstyle{plain}

\theoremstyle{definition}

\theoremstyle{remark}

\newcommand{\rqone}{how do developers apply design processes and design methods hackathons?}

\begin{document}


\title{The Developers' Design Thinking Toolbox in Hackathons: A Study on the Recurring Design Methods in Software Development Marathons}

\author{
\name{Kiev Gama\textsuperscript{a}\thanks{CONTACT Kiev Gama.  Email: kiev@cin.ufpe.br ORCID https://orcid.org/0000-0003-1508-6196},  George Valença\textsuperscript{b},  Pedro Alessio\textsuperscript{a},  Rafael Formiga\textsuperscript{a},  André Neves\textsuperscript{a},  Nycolas Lacerda\textsuperscript{b}}
\affil{\textsuperscript{a} Universidade Federal de Pernambuco (UFPE),  Recife,  Brazil; \textsuperscript{b} Universidade Federal Rural de Pernambuco (UFRPE),  Recife,  Brazil}
}

\maketitle

\begin{abstract}
Hackathons are time-bounded collaborative events that typically take between 1 to 3 days of intense teamwork to build prototypes usually in the form of software, aiming to specific challenges proposed by the organizers. These events became a widespread practice in the IT industry, universities and many other scenarios, as a result of a growing open-innovation trend in the last decade. Since the main deliverable of these events is a demonstrable version of an idea, such as early hardware or software prototypes, the short time frame requires participants to quickly understand the proposed challenge or even identify issues related to a given domain. To create solutions, teams follow an ad-hoc but effective design approach, that many times seems informal since the background of the participants is rather centered on technical aspects (e.g., web and mobile programming) and does not involve any training in  Design Thinking. 

To understand this creative process, we conducted a set of 37 interviews with people from 16 countries, consisting of 32 hackathons winners and 5 hackathon organizers. We aimed to identify the design practices (i.e., processes and recurring design methods) that are applied by winners in these types of events. Also, we conducted a focus group with 8 people experienced in hackathons (participants and organizers) to discuss our findings. Our analysis revealed that although hackathon winners with IT background  have no formal training on Design Thinking,  they are aware of many design methods and typically follow a sequence of phases that involve divergent and convergent thinking to explore the problem space and propose alternatives in a solution space, which is the rationale behind Design Thinking. We derived a set of recommendations based on design strategies that seem to lead to successful hackathon participation. These recommendations can also be useful to organizers who intend to enhance the experience of newcomers in hackathons.  

\end{abstract}


\section{Introduction}
Hackathons are collaborative events that engage people in small teams to produce software in a limited amount of time, typically in a non-stop format lasting from 1 to 3 days \citep{komssi2014hackathons}. This type of event is a growing phenomenon that emerged as a way to foster innovation and became part of the routine in tech companies and universities~\citep{warnerguo2017,pe2018designing}. Hackathons became popular worldwide in multiple contexts, being adopted for diverse purposes, such as product development~\citep{saravi2018, pe2018designing}, innovation ~\citep{angelidis2016hackathon, rosell2014}, recruitment \citep{tapia2020educational,richard2015stitchfest}, civic issues (i.e., civic hacking)~\citep{johnson2014,lodato2015issueoriented}, informal learning~\citep{nandi2016hackathons,warnerguo2017}, higher education~\citep{porras2018hackathons,gama2018hackathons} and domain-specific skills training~\citep{silver2016healthcare,wyngaard2017hacking}. 

The main deliverable in a hackathon is a demonstrable version of an idea \citep{komssi2014hackathons}, often in the form of a fully functioning software prototype \citep{pe2018designing}. Hackathons can be classified as tech-centric, much focused on exploring technologies and APIs, or focus-centric, when they are more oriented to a social issue or business objective~\citep{briscoe2014digital}. In this latter format, which is the preferred one in many disciplines~\citep{angelidis2016hackathon,zapico2013hacking, wyngaard2017hacking} and in innovation circles~\citep{flores2018can,avalos2017,rosell2014}, participants have to rapidly understand the context of the specific issue or challenge to be tackled. There is a very short amount of time to speculate about the market or user needs and propose innovative solutions addressing the event goals.  However,  developers who participate in these events do not arrive having any previous training on the design skills necessary to conceive such solutions.




The prevalent target audience of hackathons, although many times multidisciplinary, is typically formed by a vast majority of people with technical backgrounds like software developers and STEM students~\citep{briscoe2014digital, zapico2013hacking, warnerguo2017}. The typical deductive-rationalist thinking style of IT engineers leads to an approach of having a given problem and -- based on logic -- deducing the right solution. It differs from a Design thinking style, which explores a problem-solution space more openly and brings a multi-perspective comprehension that helps to deal with the ambiguity of problems~\citep{lindberg2011design}. In that sense, hackathons fit the latter style by employing an ad-hoc design approach, being more organic and oscillating from improvisation to specificity in the creation process of technical design artifacts \citep{lodato2015issueoriented}. Such an approach is aligned with the Design discipline as a whole, in which much design activity (especially what concerns conceptual design) is unplanned, intuitive and ad-hoc \citep{cross2011design}. Hackathons are a venue where participants can explore this creative approach to develop or enhance problem-solving skills~\citep{nandi2016hackathons,taylor2018everybody}.

Our experience in the organization of over 30 hackathons led us to observe a pattern in the way successful teams (i.e., winning teams) tackle the challenges. It fits the Design Thinking style, which contrasts to the deductive-rationalist thinking style of average teams that do not think as creatively (i.e., "outside of the box") and does not have such high profile results. In these hackathons, although design techniques or methods (e.g., brainstorming,  personas)  were rarely presented to participants, their strategy rather followed many Design Thinking \citep{brown,cross2011design,pe2018designing} principles towards a more user-centered and innovative approach to problem-solving, even though their educational background on STEM (Science, Technology, Engineering and Math) does not involve the learning of such skills. In hackathons, the radical collocation of participating teams helps quickly advance technical work \citep{trainer2016hackathon}. The way of thinking design might be affected by group behavior and it can also influence the thinking of other members of that group \citep{lawson2006designers}. Therefore, we believe that the way those successful teams intensely collaborate and think design as a group helps to spread knowledge on Design Thinking. These developers share a sort of Design Thinking toolbox that was not formally taught to them, but rather organically acquired through a learn-by-doing approach.



Very little has been reported about details on the usage of design methods or techniques put in practice by participants during hackathons. Many websites and blogs report on hackathons organization\footnote{Major League Hacking's guide - https://static.mlh.io/docs/mlh-member-event-guidelines.pdf} provide general-purpose guidelines\footnote{The Hack Day Manifesto - http://hackdaymanifesto.com/} for properly planning such events but without much detail on the design processes or methods involved. A recent literature review~\citep{flus2021design} sheds light on opportunities for design research in hackathons, by introducing a broad perspective indicating that hackathon participants follow the typical divergence–convergence patterns in their design process, thus confirming our field observations of an inherent Design Thinking approach in these events. However, in general, details on the ideation process used in these events are still missing in literature \citep{angarita2020we}. We can find isolated contributions about the hackathon process, under a broader view of its phases \citep{komssi2014hackathons, pe2018designing, valenca2019}. Few of these works slightly approach the design perspective of individual hackathons, such as a self-reported process in one hackathon \citep{olesen2018four} and a detailed study that focuses on a specific ideation method typically used in hackathons that used two events as sample \citep{filippova2017diversity}.

As a way to bring a contribution to literature, by going deeper into the details of the design processes and methods used in hackathons, we tried to better understand the step-by-step performed by participants of hackathons, from problem to solution. We conducted a case study of focus-centric hackathons that have  competitive focus (i.e., awards prizes) and collected data via semi-structured interviews with 32 winners and 5 organizers of hackathons from 16 different countries. In this paper, we used the phases of the Double Diamond design process model to frame the different design methods we identified as being recurring in hackathon winning teams.  We believe this set of practices help to understand the proposed challenges, create software prototypes and succeed at the event. Based on such findings, we derived a group of recommendations, also reflecting current discussions in the maturing hackathons literature.

In Section 2, we present the conceptual background that provided the foundation for this research, as well as related work. Section 3 describes our research methodology, based on a qualitative paradigm. Section 4 describes our main findings around the typical design methods adopted by hackathon winnners. In Section 5, we bring a discussion on the findings and derive recommendations based on a critical analysis of the design practices employed by the teams. Finally, in Section 6, we conclude this paper by highlighting its main contributions to academia and industry. Besides, we detail threats to its validity and future studies.

\section{Conceptual Background}

\subsection{Design Methods and Design Thinking Process Models}
\label{DT_process}

 In the Human-computer Interaction (HCI) field,  several techniques and methods have been developed, aiming to be applied to achieve the development of
usable systems for people~\citep{ogunyemi2019systematic}. There is a multitude of design methods that have been developed and documented~\citep{ogunyemi2019systematic,mueller2018handbook,hanington2012universal}, such as personas, mind mapping, Wizard of Oz, among many others. The terms ``techniques'' and ``methods'' are used interchangeably by HCI researchers in the literature because of the similar meanings of the two terms~\citep{ogunyemi2019systematic}.

Going beyond universal design methods, design thinking research started with the interest of identifying mental strategies of designers, but later its concept was stretched to a thinking process for conceiving products/services and introducing design methods into fields such as business innovation  \citep{tschimmel2012design}. Design thinking brings design practice among people without a scholarly background in design, and important process models have emerged in that field, with well-defined phases that have specific objectives~\citep{gronman2021process}. Proposers of Design Thinking process models such as IDEO's 3I model (Inspiration-Ideation-Implementation), the Hasso-Plattner Institute model (Understand, Observe, Point-of-view, Ideate, Prototype, Test) and the British Council's Double Diamond Model defend their approaches as the most appropriate ones for innovation \citep{tschimmel2012design}. Synthesizing such a process into a fixed number of steps can be very restrictive, as design thinking is an inherently iterative and usually non-linear approach. However, models help to give a comprehensible view and act as a support tool on the general steps that can be taken in a project.

\begin{figure}[t!]
\centerline{\includegraphics[scale=0.5]{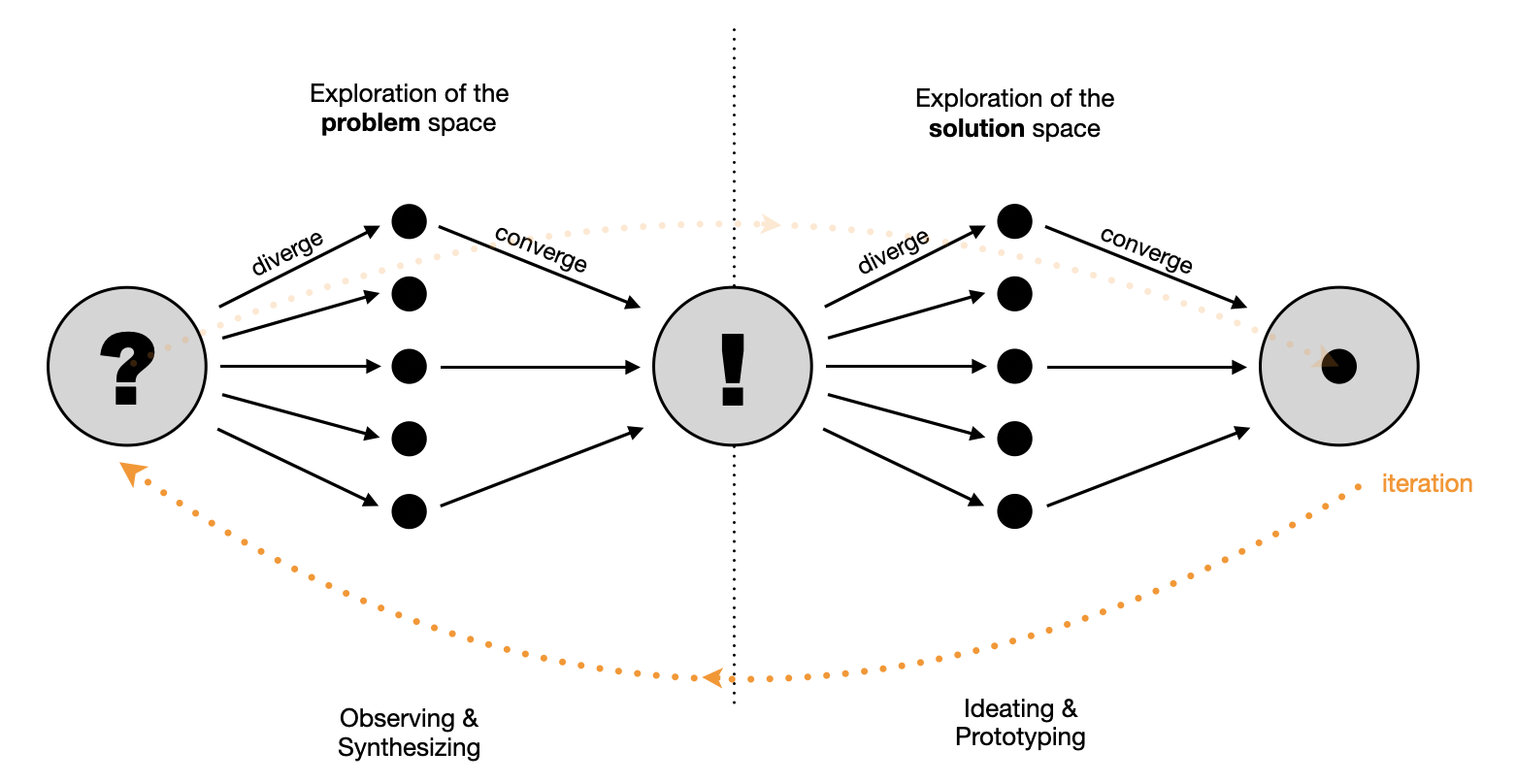}}
\caption{Exploration of problem space and solution space follow a pattern of divergent thinking followed by convergent thinking (adapted from the work of Lindberg et al. \citep{lindberg2011design}).}
\label{fig:problem_solution}
\end{figure}

Behind these models, there are key shared concepts like divergent and convergent thinking \citep{brenner2016design}, and the notion of problem and solution space, which form
a general perspective of the design process, as presented by Lindberg et al. \citep{lindberg2011design} (Figure~\ref{fig:problem_solution}). The purpose of \textit{divergent thinking} is to create choices, while \textit{convergent thinking} focuses on eliminating options and making choices \citep{brown}. The \textit{problem space} and \textit{solution space}, illustrated on Figure~\ref{fig:problem_solution}, are also commonplace in design. The goal is exploring the problem space to build a shared understanding of the problem before the actual development process starts, and exploring the solution space to promote a creative ideation process to choose the most viable path among the many alternatives generated \citep{lindberg2011design}.

\begin{figure}[t!]
\centerline{\includegraphics[scale=0.75]{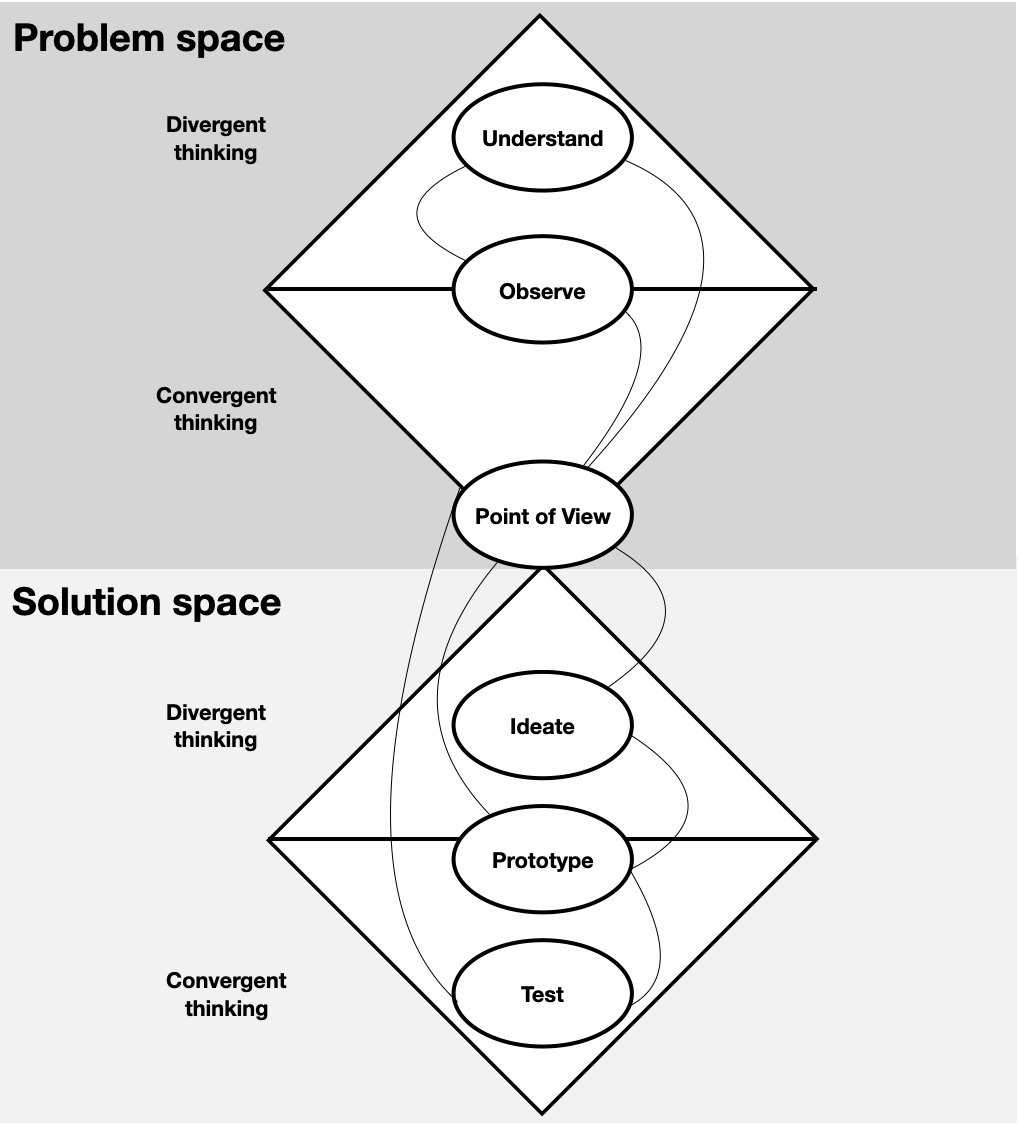}}
\caption{Merged vision of different design thinking perspectives encompassing the Double Diamond \citep{doublediamond}, the Hasso Plattner Model \citep{hassoplattner} adapted from \citep{mueller2018handbook} }
\label{fig:merged_processes}
\end{figure}

Mueller-Roterberg \citep{mueller2018handbook} provides a perspective (Figure~\ref{fig:merged_processes}) encompassing both Lindberg's perspective and the Hasso-Plattner Model, as well as the Double Diamond framework for innovation \citep{doublediamond}. The author highlights that, even though some Design Thinking processes are represented sequentially, they are inherently iterative, allowing feedback to previous phases. 

The Double Diamond Model \citep{doublediamond} indicates this divergence-convergence happening twice (the two diamonds on Figure~\ref{fig:merged_processes}), which is in compliance with Lindberg et al. \citep{lindberg2011design}. Each of the two diamonds brings the notion of exploring an issue more widely or deeply (divergent thinking) followed by focused action (convergent thinking). The first diamond exploring stakeholder needs and finding a problem definition, while the second one aims to generate multiple potential solutions and implement one of them. The four phases are depicted as follows:  

\begin{itemize}
    \item \textbf{Discover}. This stage consists of divergent thinking. The focus is on talking to stakeholders and understanding, rather than simply assuming, what the problem is.
    \item \textbf{Define.} This step converges the insights and information from the discovery phase and helps to define the problem or challenge.
    \item \textbf{Develop.} This is the first stage of the second diamond, where the focus is again on divergent thinking, generating different answers or solutions to the clearly defined problem are sought. 
    \item \textbf{Deliver.} This stage is convergent thinking and involve testing different solutions at a small-scale, eliminating what would not work and improving and choosing what would work.
\end{itemize}

\subsection{Hackathons}
In a nutshell, we can describe hackathons as time-bounded collaborative events in the format of "programming marathons" focused on solving a problem in a short time frame (1 to 3 days) usually delivering a software prototype as a result \citep{briscoe2014digital, komssi2014hackathons}. These events started in the 2000s as a method that focuses to create solutions to tackle specific challenges \citep{pe2018designing}. There are different categorizations and taxonomies on the types and formats of hackathons~\citep{briscoe2014digital,disalvo2014building,kollwitz2019hack}. We will refer to the one from Briscoe and Mulligan~\citep{briscoe2014digital} which loosely groups hackathons as tech-centric or focus-centric. The former comprises events focused on developing software using a specific technology (e.g., a hackathon aiming to promote the usage of an API). The latter involves creating software prototypes to address a specific social issue or business objective; for instance improving city transit systems. For a contextualization on our perspective on the topic, throughout this section we situate Briscoe and Mulligan's classification in regards to three main categories of hackathons we highlight: corporate, educational, and civic hackathons \citep{valenca2019}. 

\textbf{Corporate hackathons} aim at broadening participation in the corporate innovation network, with participants typically motivated by learning and networking~\citep{pe2020corporate, flores2018can}. They are commonly adopted by IT companies of all sizes, which integrate these events into their research and development activities. The resulting outcome of such collaborative events are new ideas, early prototypes, and even business plans \citep{pe2018designing}. From a perspective of the participating audience (i.e., the "crowd"), corporate hackathons can be either internal or external to an organization. Once \textbf{internal}, the hackathon stimulates the creative thinking of the organization staff to reflect on those challenges and raise new ideas. For instance, Facebook systematically runs hackathons and reports on further developing and shipping
60\% of the topics were proposed in these events \citep{raatikainen2013}. Many of these internal hackathons are tech-centric. On the other hand, \textbf{external} hackathons are open to participants outside the organization. This approach follows the open-innovation paradigm by introducing external resources in the process of crafting new solutions, which would tend to be focus-centric hackathons. In the particular case of IT companies, external hackathons can alternate between a tech-centric or focus-centric approach, or even combine the two, since in that context these events represent a strategy to support ecosystem evolution by offering a software platform for third parties to develop new products or services, and encouraging such outsiders to become complementors in the network. This is the case of well-known ecosystems such as Google’s Android and Apple’s iOS \citep{valenca2019}. In the corporate world, besides promoting platforms and technology~\citep{smolander2020acm} hackathons also are a way to attract and build a community of experts \citep{granados2019collaborative}, which help to foster a broader innovation ecosystem. There are many hackathons focused on innovation, such as TechCrunch Disrupt\footnote{techcrunch.com/tag/disrupt/}, Junction\footnote{www.hackjunction.com}, and many others that are usually listed in hackathon portals like DevPost\footnote{www.devpost.com}. These events have multiple corporate sponsors promoting their brands and products, and sometimes venture capital investors attracted by the potential of the outcomes. Thes are also under the umbrella of corporate hackathons, with events being many times a combination of tech-centric (e.g., highly focused on innovation around products and APIs) and focus-centric (e.g., specific problems to be addressed with technology, focus on business models) characteristics. 

\textbf{Educational hackathons}~\citep{nandi2016hackathons, gama2018hackathons, porras2018hackathons} are performed in association with teaching and learning activities, either as an exclusive initiative of a professor or as cooperation between academia and industry -- which is sponsoring the event. These are typically tech-centric events. In IT or Design courses, for instance, the hackathons become a contest for graduating students to address real-life issues --  thus bringing also a focus-centric approach -- in a fun scenario that enables them to intensively collaborate and enhance their abilities \citep{porras2018hackathons}. The hackathons to address civic issues, organized by the public sector or by non-governmental organizations, can be classified as \textbf{civic hackathons} \citep{johnson2014}, which focus on more socially-oriented innovation~\citep{briscoe2014digital,disalvo2014building}. These are normally focus-centric hackathons, although Government institutions also have been using such events to generate value from open data and APIs (a more tech-centric perspective), which are explored by different players (e.g., citizens, different types of companies, universities, etc.). In general, these contests leverage the idea of government as a platform~\citep{safarov2017}.

\subsection {Related Work}

The rapid pace of hackathons leaves no space for use case definition and architectural design ~\citep{thomer2016}, but these events hold significant potential to explore users' or customers’ interactions and rapid identification of their needs. Hence, hackathons can benefit from design thinking frameworks to produce better and more feasible customer-oriented solutions~\citep{saravi2018}. Some organizations perceive the importance of design methods, including a specific preliminary phase in the hackathon to foster participants' ideation. For instance, to promote innovation some hackathons can start with activities outlining the basics of Design Thinking methodologies and skills to participants~\citep{avalos2017}. The basic training enables groups to focus on key suggested methods to address the challenges, from artifacts to identify users and sponsors to artifacts to validate the ideas.

Overall, hackathons can have important learning outcomes for participants in regards to design thinking.  In a project described by~\cite{lyndon2018hacking}, high-school students participated in a hackathon where they had to create an effective user interface for a mobile application in the medical domain, to support surgical capacity inventory. Hands-on activities in the first day helped to build the necessary skills for the prototyping phase that took place in the second day. According to what students positively reported, they developed more empathy toward users and got acquainted with a design process to develop a product. The work from Suominen et al.~\citep{suominen2018educational} describes an educational hackathon in the context of university-industry collaboration in which teachers introduced students to design thinking methods. Page et al.~\citep{page2016use} report on a corporate-sponsored 4-day hackathon focused on design education that was performed at a university.

Although there is evidence of design thinking being addressed in hackathon literature, empirical studies on such design practices in the context of those events are considered scarce~\citep{angarita2020we}. The design thinking process or processes inherent to hackathons -- either via formal training or as practices typically adopted by participants -- are rarely reported in detail. In fact, the literature often focuses on broader views with proposals for organizing the whole event, such as the adaptation of a Lean Innovation Model for a 2-day hackathon~\citep{flores2018can}. Another example is the work from Frey and Luks~\citep{frey2016innovation}, who depict different phases of a hackathon and enumerate activities expected for each phase on what they called the \textit{innovation-driven hackathon pattern}. In the first phase (Problem), the user needs are the priority. The second phase (Solutions Alternatives) aims at thinking about solutions for the identified problems; followed by a third phase (Prototypes) showing a prototype with the key features (i.e., an MVP) of the alternatives. However, that work does not explicitly mention any specific design method.   \cite{poncette2020hackathons} briefly details what the called “hackathon flow”: brainstorming sessions using post-its and diagramming; prototyping (software mockup); and a pitch to present the working prototype. ~\citep{suominen2018educational} lists the design techniques or tools (brainstorming, mindmapping, etc) that were taught in the studied hackathon, but the authors did not offer additional details of their use nor interpreted such practices. ~\citep{page2016use} presents a 4-day event was structured upon the Double Diamond process model, the descriptive analysis of the authors was not focused on design issues, but rather on general aspects or characteristics of a hackathon such as icebreakers, leadership and team roles, and workplace environments. 
McGowan~\citep{mcgowan2019role} presents a model for supporting the construction of services by librarians in hackathons, showing clear phases based on design thinking: Assess,
Define Objectives, Develop, Deliver, and Measure. However, the report does not focus on what design methods are applied in each phase.

We associate the limitations of details around design methods in hackathons due to the informal and unstructured nature of the design steps, which are hard to identify in practice. Another assumption for such lack of studies is the typical focus of hackathon organizers on technical concerns, as highlighted in~\cite{frey2016innovation}, instead of focusing on design activities. The recent literature analysis by Flus and Hurst~\citep{flus2021design} brings a broad perspective on overall design practices by hackathon participants, highlighting the Double Diamond design process as a model that encompasses the typical divergence–convergence patterns that hackathon participants use. However, the article does not provide any details on which design methods or creativity techniques can be typically followed through the phases of the double diamond. Our main contribution in this work lies in not just identifying and listing such methods but also in analyzing how these design methods are being adopted and applied in hackathons. By discussing the hackathon design process with experienced participants and organizers, identifying the recurring methods and representing a perspective of an overall process through phases, we generated recommendations to improve such a roadmap.

\section{Methodology}
\begin{figure}[b!]
\centerline{\includegraphics[scale=0.45]{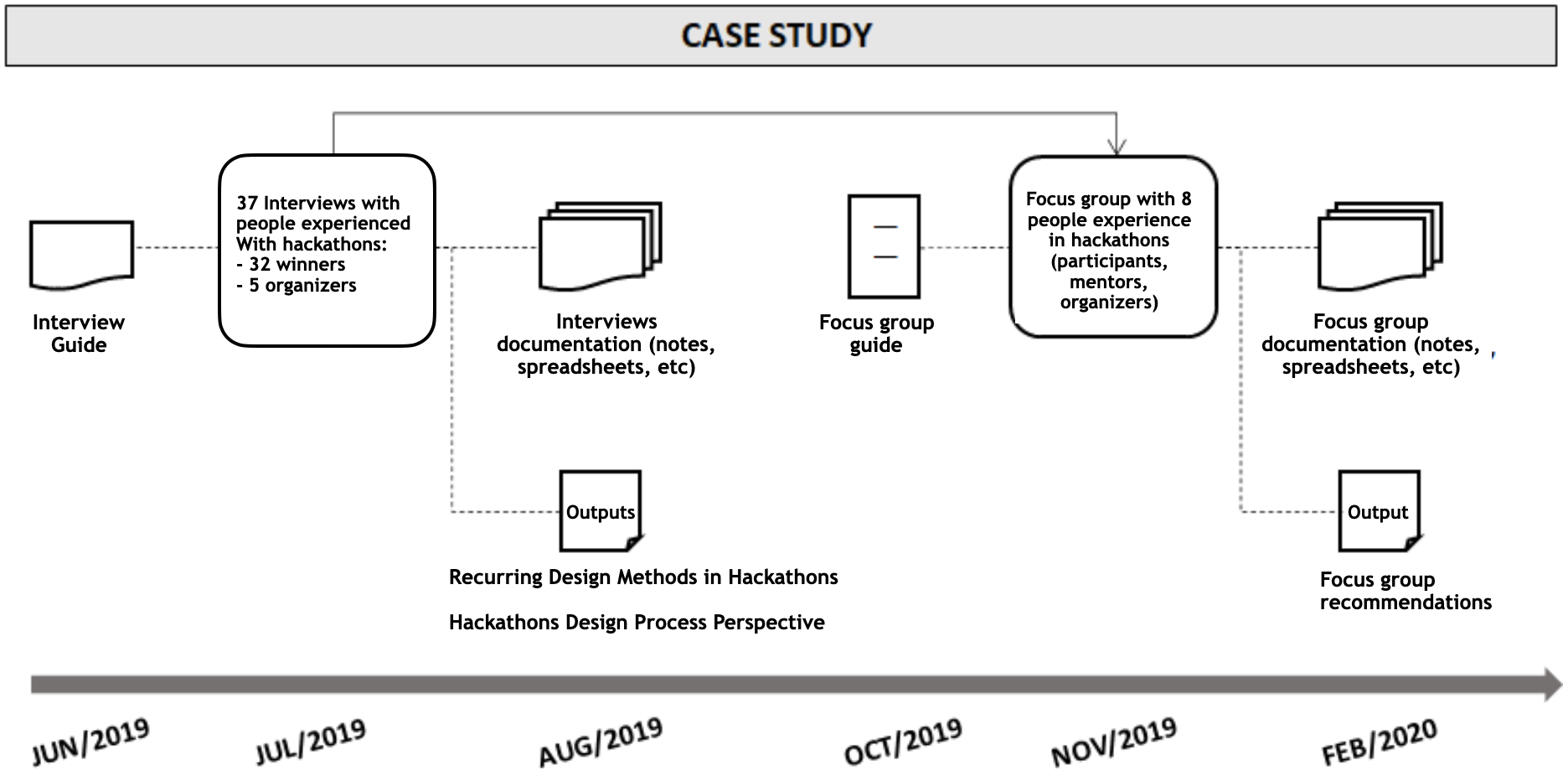}}
\caption{Two main research phases.}
\label{fig:phases}
\end{figure}

This research followed a qualitative approach to answer the following research question (RQ): \textit{\rqone{}} These design methods inherent to competitive and focus-centric hackathons were the unit of analysis of a \textbf{descriptive case study} that we conducted with participants and organizers of these events. This type of case study sets to describe a phenomenon within the data collected \citep{zainal2007case}. 

To understand our unit of analysis, we performed a wide set of \textbf{interviews} as well as a \textbf{focus groups}. Such empirical data paved the way for a \textit{inductive} research approach when we described the phenomenon in its particular context and obtained insights. After identifying patterns within data, we adopted an \textit{abductive} approach to transcend collected data and introduce a creative aspect in data analysis \citep{sjoberg2008}. Such abductive reasoning presents a conceptual explanation by analogy with relevant ideas in domains that are well-understood \citep{haig2018}. Hence, to draw one of the possible representations of the design processes followed in hackathons, we combined established concepts from Design with concepts extracted from the state of the practice in hackathons.

The following sections explain our data collection procedure, together with the associated analysis and results, as represented in Figure~\ref{fig:phases}.

\subsection{Interviews}

In June 2019, we structured the interview guide based on discussions involving the authors with a design background, about a typical Design Thinking cycle. The author with the largest experience in conducting hackathons has a Computer Science background -- with some knowledge on Design Thinking although no formal training on the theme -- and presented the study hypothesis to the authors from the Design field, who had a formal background on the topic but little experience in conducting hackathons. The hypothesis on the hackathon participants' informal knowledge of design methods was based on non-systematic observations made as part of the accumulated experience in the organization and conduction of hackathons of different classifications (focus-centric, tech-centric and a combination of these two) and types (corporate, civic and educational). This led to rounds of discussions that helped reduce the initial bias in questions asking specific design techniques, which were exchanged to more general questions, such as trying to describe the step-by-step process, and broader questions about higher-level design activities (e.g., generation and selection of alternatives, prototyping, testing). 

This process resulted in the interview guide that was divided into two main parts: the first part consisted of demographic questions about the interviewee (e.g., age, educational degree), while the second one involved specific questions about the hackathons dynamics, involving teamwork and the design process. We started the second part by asking interviewees about their groups in the hackathon, addressing topics such as the process of team formation and team composition (e.g., the roles played by each participant). The subsequent questions explored the steps of the design process executed during the hackathon. 
In total, we included 21 questions in our protocol, which is available in the appendix. 
We must highlight our adoption of semi-structured interviews to ensure that a single basic structure was followed, but new questions could be brought up based on interviewees discourse as a means to obtain a more in-depth understanding of the topics. In the case of organizers, we adapted the way questions were asked by bringing "do you propose..." or "do you see any occurrence of..." types of inquiry.

We selected the interview participants considering their experience in competitive hackathons that were corporate-sponsored -- but not necessarily events with a single sponsor -- and more centered on problems (i.e., focus-centric), with specific challenges where innovative solutions are mostly appreciated. This would allow us to understand what is the participant's path to find such innovations. Another fundamental selection criteria concerned the participant's experience in winning a prize (1st-3rd place) in a hackathon at least once. We chose to contact winners for two reasons: (1) from our experience in organizing these events we notice the combination of design methods as a common pattern in winning teams while other participants were very tech-centric and less worried about the design process; and (2) it is difficult to have access to the list of participants in hackathons websites, where typically only the winners are announced. All hackathon winners we contacted were evaluated by a jury of experts in the events they won. Interviewees were contacted through professional social networks and referrals. {In the former strategy we found participants by contacting people through their LinkedIn or GitHub profiles after finding them in searches for event winners on hackathon portals such as DevPost and Web sites of large hackathons such as Junction. The latter one consisted of asking peers from the authors' professional networks to indicate experienced hackathon participants who have already won prizes in such events. In both cases, to avoid participant bias, only people unknown to the authors have been contacted. Thus, we avoided reaching participants from events we previously organized.} In July 2019, we performed a total of 37 interviews through video call software with people experienced with hackathons: 5 organizers from varied countries (Australia, Canada, Brazil, France and Portugal); and 32 winners of hackathons from 16 countries (including winners of the largest hackathons in Europe, Singapore and Silicon Valley). They all have a software development background, including interviewee P22, who has a Bachelor of Arts in Architecture but is a front end developer. The complete profile of the hackathon organizers is presented in Table \ref{table:organizers}, while the profile of the participants is shown in Table \ref{table:participants}. 

\begin{table}[!h]
\begin{tabular}{llll}
\textbf{Participant} & \textbf{Country} & \textbf{Gender} & \textbf{Company's main activity} \\
\hline
O1                   & Canada           & Male            & Banking                          \\
O2                   & Australia        & Male            & Hackathons Organization          \\
O3                   & Portugal         & Male            & Hackathons Organization          \\
O4                   & Brazil           & Male            & Hackathons Organization          \\
O5                   & France           & Male            & Hackathons Organization         
\end{tabular}
\caption{Profile of the interviewed hackathon organizers. The first one is responsible for organizing hackathons in a banking institution while the others are part of institutions focused on organizing hackathons.}
\label{table:organizers}
\end{table}

\subsection{Data Analysis}
The duration of the interviews ranged from 8 to 52 minutes, with an average duration of 25 minutes. In total, we obtained 15 hours and 12 minutes of recordings. To analyze this rich dataset, we adopted a live coding procedure, which means "not transcribing everything" \citep{parameswaran2019live}. Based on this approach, we could code the interviews "on the fly”, writing down memos and extracting relevant quotes that were tagged with the respective timestamps where they show up in the digital audio file.
Researchers from a design background and experience in applying design methodologies analyzed the whole dataset to identify occurrences of design methods in interview transcripts. Each audio file was listened to separately by two researchers: one who generated analytical memos \citep{saldana2013coding} and registering relevant quotes when evidence of a design method was found on the interviewee's speech and another who listened to the audio to confirm the identification of the techniques and eventually other techniques. The second researcher also complemented the data generating additional memos or extracting more quotes. Such selective transcription procedure was supported by notes taken in an interview diary created by the researchers while performing the interviews. We also created a spreadsheet to describe the profile of each participant, which were also based on analytical memos that detailed the researcher perspective about the interviewee and highlighted some characteristics perceived (e.g., tech-focused, design-focused). In Table \ref{table:participants}, we present the main characteristics of the interviewees (i.e., country, age, gender, experience with software development, background, total of participation in hackathons, and number of prizes obtained).

As a complementary step to examining the interviewee's discourse, the group of researchers analyzing the interviews did two face-to-face meetings to reach a consensus on the common characteristics of the design processes followed by interviewees and the individual techniques that were mapped in the analysis. We could also extract and map a group of best practices on the use of such design and creativity methods. This analysis process was performed in August 2019 and revealed we had already reached saturation close to the 30th interview, with no new occurrence of design methods on the subsequent interviews that were analyzed. This ensured us that no additional interviews would be necessary. As a main result of the analysis, we enumerated the recurring design methods that hackathon winners apply,  as presented in Section~\ref{sect:results}. We framed those methods under the phases of a Design Thinking  process model that represents well the short cycle of a hackathon~\citep{flus2021design}, although we do not intend to limit a hackathon design process to be restricted to such model.

\begin{table}[!h]\footnotesize
\begin{tabular}{lllllrr}
            &           &     &        &                 & \multicolumn{2}{l}{\textbf{Hackathon Experience}} \\
\textbf{Participant} & \textbf{Country}   & \textbf{Age} & \textbf{Gender} & \textbf{Background*}      & \textbf{Participation}          & \textbf{Prizes}          \\
\hline
P1          & Angola    & 25  & Male   & CS student      & 1                      & 1               \\
P2          & Argentina & 28  & Male   & BSc. CS         & 30                     & 5               \\
P3          & Australia & 23  & Male   & BSc. CS         & 5                      & 2               \\
P4          & Australia & 26  & Male   & BSc. SE         & 4                      & 1               \\
P5          & Australia & 35  & Male   & BSc. CS         & 5                      & 3               \\
P6          & Brazil    & 21  & Male   & CE student      & 6                      & 4               \\
P7          & Brazil    & 19  & Male   & CE student      & 4                      & 1               \\
P8          & Brazil    & 23  & Female & CS student      & 2                      & 1               \\
P9          & Brazil    & 22  & Male   & CS dropout      & 22                     & 16              \\
P10         & Brazil    & 24  & Male   & BSc. IT         & 6                      & 2               \\
P11         & Brazil    & 31  & Male   & BSc. CS         & 4                      & 1               \\
P12         & Canada    & 18  & Male   & CS student      & 20                     & 3               \\
P13         & Egypt     & 21  & Male   & BSc. CE         & 4                      & 4               \\
P14         & England   & 24  & Male   & BSc. IT         & 4                      & 1               \\
P15         & Finland   & 25  & Male   & BSc. IT         & 10                     & 6               \\
P16         & France    & 22  & Male   & CE student      & 2                      & 2               \\
P17         & France    & 23  & Male   & CE student      & 1                      & 1               \\
P18         & France    & 22  & Male   & CE student      & 4                      & 3               \\
P19         & France    & 21  & Male   & CE student      & 2                      & 1               \\
P20         & France    & 21  & Male   & IS student      & 2                      & 1               \\
P21         & India     & 19  & Male   & CE student      & 3                      & 1               \\
P22         & Japan     & 25  & Male   & BA Arch         & 3                      & 2               \\
P23         & Mexico    & 23  & Male   & CE student      & 4                      & 2               \\
P24         & Mexico    & 25  & Male   & BSc. ME         & 2                      & 2               \\
P25         & Peru      & 45  & Male   & BSc. CS \& Math & 100+                   & 30+             \\
P26         & Peru      & 38  & Male   & BSc. IT         & 5                      & 1               \\
P27         & Peru      & 23  & Male   & BSc. IE         & 14                     & 6               \\
P28         & Portugal  & 28  & Male   & BSc. CE         & 3                      & 1               \\
P29         & Singapore & 18  & Male   & CS student      & 7                      & 3               \\
P30         & USA       & 21  & Male   & CS student      & 4                      & 1               \\
P31         & USA       & 21  & Male   & CS student      & 6                      & 5               \\
P32         & USA       & 33  & Male   & BSc. CS         & 4                      & 1              
\end{tabular}
\caption{Profile and distinct backgrounds  of interviewed hackathon participants who won different events evaluated by expert juries. \\
* Arch = Architecture, CE = Computer Engineering, CS = Computer Science, IE = Industrial Engineering, IS = Information Systems, IT = Information Technology, ME = Mechatronic Engineering}
\label{table:participants}
\end{table}

\subsection{Focus Group}

To enrich the resultant analysis with additional perspectives and new insights, we decided to perform a focus group with people experienced with hackathons. We planned this step between October and November 2019. In total, we selected 8 people, with a background on Design and IT-related courses, as described in Table \ref{table:focus_group}. Most of them (5) had previously acted as organizers and participants of hackathons, which enabled them to present opinions from both perspectives. {These people were participants of previous hackathons we organized and people from the authors' professional networks who have experience in participating or organizing that sort of event.}

We started the session by presenting key concepts (i.e., focus-centric hackathons, and design thinking's goals and models) to form a common conceptual framework. Then, we described the goals and rules of the focus group, which enabled participants to understand the dynamics envisaged for the session. Initially, we presented a representation of the design process that we mapped in hackathons, as a result of the interviews previously performed. The participants were asked to critically analyze the process considering their experience with hackathons. For instance, assessing whether they would make changes such as adding, removing or moving methods to other phases. Finally, we promoted a discussion about (i) benefits brought by the design methods typically applied in the design process that we represented as the Double Diamond  and (ii) overall recommendations for hackathon participants from a design perspective.

The audio recording of about an hour was also transcribed via live coding, with codes such as \textit{how to use the design methods}, and \textit{best practices for a hackathon}. The transcript was discussed among the researchers to interpret quotes from participants and clarify specific aspects raised during the session. The resultant data interpretation provided us with fine-grained suggestions on how to apply the design methods as well as practices to reinforce or avoid in a hackathon. Such analysis generated a set of recommendations, which we present in Section
~\ref{sect:results}.

\begin{table}[]\footnotesize
\begin{tabular}{llllrr}
\textbf{}    & \textbf{}       & \textbf{}           & \textbf{}              & \multicolumn{2}{l}{\textbf{Experience in hackathons}} \\
\textbf{Age} & \textbf{Gender} & \textbf{Background} & \textbf{Roles}         & \textbf{Participation}      & \textbf{Mentoring}      \\
\hline
21           & Male            & CE student          & Dev/Business/Design    & 8                           & 1                       \\
41           & Female          & PhD Design          & Hackathon organization & 30+                         & 20+                     \\
22           & Male            & CE student          & Dev/Business           & 12                          & 4                       \\
23           & Female          & BA Design           & Design/Business        & 3                           & 0                       \\
25           & Male            & EE student          & Dev                    & 9                           & 2                       \\
23           & Male            & EE student          & Dev                    & 5                           & 0                       \\
22           & Male            & CE student          & Dev/Business           & 12                          & 3                       \\
28           & Male            & Design student      & Design/Management      & 3                           & 0                      
\end{tabular}
\caption{Participants of the focus group and their background (CE = Computer Engineering; EE = Electronic Engineering)}
\label{table:focus_group}
\end{table}








\section{Design Methods used in Hackathons}
\label{sect:results}
In this section we start with a subsection focused on the perspective from organizers, followed by a subsection with the participants' perspective on the design process. In the last subsection we take into account the opinions of the focus group participants, suggesting complementary design methods throughout the phases of the process.

\subsection{Organizers Perspective}
Before bringing a perspective of participants, we wanted to understand how organizers address the notion of activities that would comprise a design process in hackathons and how they see some practices of the participant. The main results we highlight in this section concern perceived practices on understanding user needs and prototyping; and attempts of organizers suggesting a process for participants.

Concerning attempts to address user needs (or, an audience's needs since there are no users sometimes), O3 perceived that short hackathons (12h to 24h) solutions were self-biased, without a strong foundation to justify decisions on the choice of a problem or feature, while in longer hackathons (48h) there are clear attempts to target a problem of a specific audience. In the case of organizer O1, who is from a banking institution, the hackathons were short events on Saturdays (from 9 AM to 6 PM), with the advantage of having stakeholders from the bank available during the whole day. The problems and solutions were clear, thanks to the stakeholders, but they did not interfere in the choices of participants. Due to the limited time, the focus was on ideas or innovative approaches rather than technologically appealing solutions.

Organizer O2 reported his institution being focused on 24-hour corporate-sponsored hackathons targeting university students. He mentioned tech groups being focused on the product/prototype itself, while multidisciplinary groups tend to go beyond that to explore business models and user research. Despite the lack of recommendation of design processes or design methods, organizer O2 observed common practices in the first hours of the hackathon, such as performing (i) a brainstorming session and designing wireframes to choose project ideas and (ii) quick surveys through online questionnaires on tools like Typeform or Google Forms to validate ideas and gather some user information. Design practices seemed more focused on UI aspects, rather than conception aspects.

We found many hackathon organizers proposing overall dynamics at the beginning of the event, which includes idea pitching to support team formation -- although often times groups come already formed. Concerning a project conception process, organizers typically do not propose any specific model or set of design methods to conduct the activities to develop the solution. The context of organizer O5 was very particular. He is in charge of a university department that takes care of innovation initiatives, so the mission is to foster a creative mindset. The hackathons they organize are more mostly oriented to idea generation and value proposition, rather than creating fully functional application prototypes or proof of concept. So, this was the only case of specific creativity techniques being put into practice.

The closest to a product-focused process we found consisted on what was reported by organizer O4. The company where he works focuses on conducting corporate-sponsored hackathons. They have general guidelines for participants, mentors and clients. Their purpose to systematize hackathons organization resulted from observations of groups failing, as illustrated below. Hence, they started promoting a Design Thinking mindset to participants.

\textit{"There were groups that did not know what problem they are tackling. They were proposing a solution without a problem. It was a sort of self-sabotage. They would stick to an idea because of the technological solution" - O4} 

This company started framing the hackathons using the Double Diamond as a framework, but not stressing the usage of too many design methods or creativity techniques as it was the case of organizer O5. Basically, the usage of brainstorming and persona in the problem space, and the recommendation of presenting an MVP (Minimum Viable Product) using storytelling in the solution space. The reported hackathon participants came from different backgrounds (IT, design, business) and, once they knew any design technique or method (e.g., user journey), they employed it as they want.

\subsection{Participants Perspective}

The majority of the winners we interviewed never had formal training in any of the identified design methods. In one of the questions, we wanted to speculate about the influence of designers in the group. According to interviewees, who were all developers, the eventual designers who would occasionally be part of the groups did not bring any creativity techniques. They were mostly focused on UI Graphical Design. User research and ideation was an activity shared by all team and sometimes led by the interviewees (IT background). 

We found much evidence of design methods being learned from other peers during hackathons or applied in a simplified or adapted manner. More important than using individual methods is having a design thinking mindset. We noticed that participants hold strong critical thinking, e.g. they generally reflect and investigate a given problem before focusing on the possible solutions. With hackathons being a sort of venue where the freedom of thinking, the ad-hoc nature of working seems to be inherent of the prevalent culture. Some people might see a process as a restraining device for creativity, as stated by the two participants who reported a bad experience with processes being imposed:

\begin{quote}
\textit{"Some hackathons have this workshop style hackathon where they actually ask you to follow their process. It kind of feels a bit forced. (...) The way I usually like to do it's very informal."}  - P15
\end{quote}

\begin{quote}
\textit{"It takes away our freedom. Each team has its own way of organizing its work. Having (to follow) very defined stages is like imposing a project structure. That doesn't leave a good feeling. Of course, in this sense, I would value if there was someone who knew or would suggest or support teams that have no idea how to organize, but should not be imposed."}~-~P26
\end{quote}

\begin{figure*}[!ht]
\caption{The Double Diamond representating a Design Thinking process in hackathons, with the main recurring design methods that are used by participants.}
\label{fig:diamond}
\centering
    \includegraphics[width=0.98\textwidth]{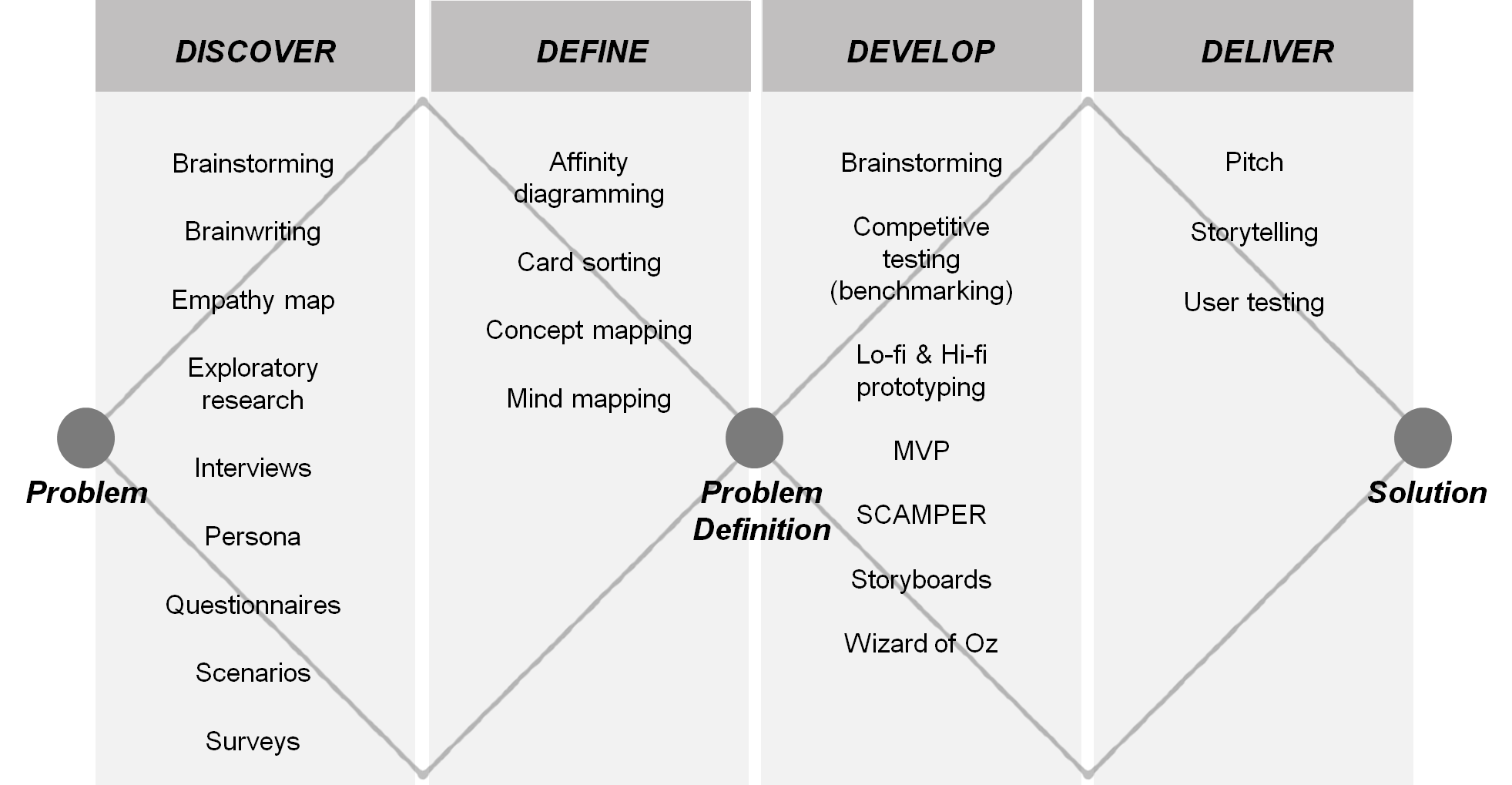}
\end{figure*}

Interviewees mentioned hackathons as not having a structured approach. However, our analysis showed that the approach these hackathon winners described follow a convergent-divergent approach that is clearly a design thinking process, where many design methods are employed typically in an informal or adapted way. In particular, such a process was similar and consistent among interviews. 
As contradictory as the above statements from P15 and P26 might sound in contrast to what the upcoming sections show, we found a general design process taking place after analyzing what the interviewees described. There are clear convergence-divergence phases followed by the hackathon winners we interviewed, characterizing a Design Thinking process in place. In Figure~\ref{fig:diamond}, we present a perspective showing the recurrent design methods placed under the different phases model of the Double Diamond design process model. Although we cannot claim that there is just one design process model for hackathons, we found the Double Diamond as a model that easily encompasses what was described by interviewees. This choice is also reinforced by recent hackathon literature \citep{flus2021design}. The Double Diamond moulds a very short design sprint, where the compact timeframe makes it less iterative and more linear. The Design Council proposes a 4-phase framework (discover-define-develop-deliver) with a set of design methods for each phase, however we only take the 4 phases and enumerate the design methods in the corresponding phases based on our analysis. In particular, a large range of methods \citep{mueller2018handbook,hanington2012universal} is suited for different phases of the design process. Hence, some of them may be applicable to more than one phase or be usable in other contexts. 

The next sections describe our mapping and understanding of the methods being used by participants of hackathons, according to our interviews. What we bring here is not a prescription of methods or imposition of the Double Diamond as the sole design process model that is used in hackathons, but rather a perspective resulting from an analysis of the state of the practice in hackathons.

\begin{table}[ht]\small
\begin{tabular}{lp{0.275\textwidth}p{0.55\textwidth}}
\textbf{Phase}                     & \textbf{Technique}                          & \textbf{Occurrence}                                                                                                                             \\
\hline
\multirow{9}{*}{\parbox[t]{2mm}{\multirow{3}{*}{\rotatebox[origin=c]{90}{Discover}}}} & Brainstorming                      & P3, P4, P5, P7, P8, P9, P10, P13, P15, P16, P17, P18, P19, P20, P21, P22, P25, P27, P29                                               \\
                          & Brainwriting                       & P7, P8, P12, P13, P28                                                                                                                 \\
                          & Empathy map                        & P23, P27                                                                                                                              \\
                          & Exploratory research               & P3, P6, P5, P13, P21, P25, P30, P31                                                                                                   \\
                          & Interviews                         & P3, P11, P16, P20, P31                                                                                                                \\
                          & Persona                            & P1, P7, P18, P20, P21, P23, P26, P27, P28                                                                                                                    \\
                          & Scenarios                          & P1, P20, P21, P28                                                                                                                     \\
                          & Surveys                            & P13, P31                                                                                                                              \\
                          & Questionnaires                     & P4, P14, P31                                                                                                                          \\
                          \hline
\multirow{3}{*}{\parbox[b]{2mm}{\multirow{3}{*}{\rotatebox[origin=t]{90}{Define}}}}   & Affinity diagramming               & P15, P17, P18, P21                                                                                                                    \\
                          & Card sorting                       & P5, P13, P31                                                                                                                     \\
                          & Concept mapping                    & P9, P14, P20, P21, P25                                                                                                                \\
                          & Mind mapping                       & P10, P16, P19, P29                                                                                                                    \\
                          \hline
\multirow{7}{*}{\parbox[t]{2mm}{\multirow{3}{*}{\rotatebox[origin=c]{90}{Develop}}}}  & Brainstorming                      & P14, P21, P23, P24, P26, P29, P32                                                                                                     \\
                          & Competitive testing (benchmarking) & P2, P3, P4, P6, P7, P8, P10, P11, P12, P13, P14, P15, P16, P17, P18, P19, P20, P21, P22, P23, P25, P26, P27, P29, P30, P31, P32       \\
                          & Lo-fi \& Hi-fi prototyping         & P3, P5, P7, P8, P9, P10, P11, P12, P13, P14, P15, P16, P17, P18, P19, P20, P21, P22, P23, P24, P25, P26, P27, P28, P29, P30, P31, P32 \\
                          & Minimum Viable Product (MVP)       & P5, P4, P7, P11, P20, P24, P30                                                                                                        \\
                          & SCAMPER                            & P5, P6, P7, P10, P13, P15, P16, P18, P19, P20, P21, P25, P26, P27, P30, P31                                                           \\
                          & Storyboards                        & P1, P2, P3, P9, P11, P13, P29                                                                                                         \\
                          & Wizard of Oz                       & P18, P19, P22                                                                                                                         \\
                          \hline
\multirow{3}{*}{\parbox[t]{2mm}{\multirow{3}{*}{\rotatebox[origin=t]{90}{Deliver}}}}  & Pitch                              & P3, P4, P6, P7, P9, P12, P13, P14, P15, P16, P17, P18, P19, P20, P21, P22, P23, P24, P26, P27, P28, P29, P30, P31, P32                                   \\
                          & Storytelling                       & P3, P15, P16, P17, P19, P21, P22, P31                                                                                                 \\
                          & User Testing                       & P2, P3, P5, P6, P8, P9, P10, P11, P13, P14, P16, P18, P21, P25, P27, P30, P31                                                        
\end{tabular}
\caption{Occurrence of the design methods identified in each interview (Participants P1-P32) }
\label{table:design_methods}
\end{table}

\subsubsection{Discover}


According to the majority of interviewees, presenting a solution to a well-grounded problem is more important than showing a complex or hi-tech application that solves no one's problem. First, they need to think about the problem, not the solution. They shared the understanding that this fact was what put them ahead of the competition in the hackathons they won when compared to other groups that did not win prizes. However, finding a problem requires an effort concerning requirements understanding. In hackathons, a key challenge is overcoming the confinement and the limited timeframe of the event in order to find a way to access potential users and try to understand their needs. Team members usually end up speculating based on their own assumptions. They also ask mentors and other participants to play the role of users. Hence, part of the understanding of the domain may come from informal discussions with domain specialist mentors. Besides,  Web searches (e.g., Google, Bing) are the typical mechanism for gaining understanding in the target domain. 

When the hackathon theme is previously made available, some participants reported doing an initial study or defining a hot topic within the big problem presented. Such research enables groups to gain time before the event. Otherwise, when the theme or challenge is revealed in the first hours of the event, participants have to rush to have a better understanding of the problem trying to contact external people or performing general Google searches as well as accessing resources such as discussion forums, Reditt, Quora, Facebook groups, etc. For instance, one of the interviewees mentioned a hackathon targeting solutions for the medical domain and chose to address Parkinson's disease. His team followed prompts to help to guide their research on that domain:  \textit{"What kind of problems do Parkinson's disease patients have? We saw there are a lot of motor skills problems"} (P30). \newline 


\noindent{\textbf{Exploratory research}}. We also noticed that many of the experienced participants we interviewed (P3, P6, P5, P13, P21, P25, P30, and. P31) performed more structured research to discover problems and user requirements, being very close to the format of exploratory research. In design literature, this is described as a method typically conducted in the earliest stages of a design process, acting as an immersive experience to develop knowledge and empathy about the target audience, encompassing other design methods, such as surveys, questionnaires, participants observation, among others \citep{hanington2012universal}. Interviewee P31 performed exploratory research in a very consistent way, combining observations and surveys via interviews and questionnaires. 

\begin{quote}
\textit{“When going to the venue, we used to see what is going wrong around us. In Los Angeles, we saw there were many blind people on the streets. So, we made a cool object recognition glasses project (…) if it is a 48h hackathon, you have time (to contact stakeholders). One or two people from the group ask a very selected number of questions and that would give you a clear-cut view. We had a lot of contacts, ranging from shop owners to people who are visually impaired. If we don’t have fruitful answers, we call them up. (…) Questionnaires on Survey Monkey or Google Forms were made available only in the first 6 hours of the hackathon.”}  - P31 \newline 
\end{quote}

\noindent{\textbf{Surveys, questionnaires and interviews}}. However, other participants with no formal training also did surveys, interviews and questionnaires. According to the many methods described by Hanington and Martin \citep{hanington2012universal}, surveys are composed of questionnaires and interviews, but these two can also be done as isolated practices. Questionnaires were mentioned by three  participants (P4, P14, and P31), but sometimes the term survey was used to describe a questionnaire, as P14 mentioned: \textit{"... we survey the required data. If we are dealing with a community, we'll put questions on the community channel"}. The survey method combining questionnaires and interviews was performed by P13 and P13. In the case of P13, for instance, he highlighted: \textit{"we do open or public surveys and ask family and friends for their opinions"}. Others relied mostly on interviews (P3, P11, P13, P16, P20, and P31). Participant P16, for instance, used data collected during this phase in the project pitch to show evidence of a well-founded problem.

\begin{quote}
\textit{"In the first hackathon, we called quite a few people. We spent a good hour calling our acquaintances. We called doctors and nurses around us. This was very important for us to develop our specifications and subsequently sell our product to the jury, supported by quotes from users." } - P16
\end{quote}


From the interviews we understood that when the identified target audience is a specific group that may be difficult to reach (e.g., visually impaired), Internet forums become a useful resource. In 48h hackathons, participants are more likely to contact people through social networks by adopting online questionnaires or to perform quick interviews with a small and improvised set of questions. The studied participants reinforced the need to avoid long and biased questionnaires or interviews, which may take too much time and generate poor results.   \newline 

\noindent{\textbf{Brainstorming}}. We observed brainstorming being applied as one of the main divergent thinking techniques during the Discover phase, as evidenced in Table~\ref{table:design_methods}. Brainstorming helps quickly enumerating relevant problems or challenges that could be potential candidates to be addressed during the hackathon. Although being commonly used for inspiration and idea generation during an ideation phase \citep{ambrose2009basics}, this method is essentially diverging -- which is the focus of the Discover phase. The brainstorming sessions described by interviewees seemed to be very informal, not necessarily having the role of a mediator as proposed in the original method \citep{osborn2012applied}. These sessions follow the principles of free expression of ideas, avoiding criticism so that group members are not inhibited to contribute. As proposed in the technique, ideas will be judged after the brainstorming session, as illustrated by P8 who described the process her group did together:
\begin{quote}
\textit{"We would see the theme (of the hackathon) and tried to brainstorm the idea that we were planning, then we select the best ones until reaching a consensus on what everyone would like to work on. It was more decentralized; everyone would write it down. In the end, we chose one (idea)"}  - P8 \newline 
\end{quote}

\noindent{\textbf{Brainwriting}}. We identified brainwriting as an alternative to brainstorming sessions, although less frequent in interviews (P7, P8, P12, P13, P28). It had variations that can be more systematic, such as the 6-3-5 technique \citep{schroeer2010supporting,rhorbach1969kreative}. Essentially, in brainwriting instead of the group writing ideas together, each participant writes their ideas alone and then the selection is done collectively. {Generally, it follows very closely what is called nominal group process \citep{van1971nominal}, which include steps such as the silent generation of ideas, group discussion and voting.} It is especially useful in groups whose members do not know each other, avoiding the imposition of dominant members that may take place in brainstorming \citep{litcanu2015brain}. In practice, some interviews showed that different adaptations of brainwriting in hackathons can be close to a collective brainstorming session: \textit{"we just have a Google Doc, so we can all just see and write at the same time"} (P12). The individual approach was better illustrated by participant P28, when his group was trying to enumerate possible topics (which he referred to as "ideas") to address in their project: 

\begin{quote}
\textit{ "We first did an individual generation of ideas. Each person wrote on a paper sheet ideas they wanted to see and then we talked and chose one of them."} - P28 \newline 
\end{quote}


\noindent{\textbf{Persona}}. Hackathon winners reinforced the relevance of understanding users’ motivations. Many participants (P1, P7, P18, P20, P21, P25, P30, and P31) cited the creation of personas \citep{cooper2004inmates} to understand the problems and needs that potential users may have. A persona is a general user archetype that can be used at the beginning of an ethnography process or in the end, as a result of user research. This method is key to avoid any tendency the developer may have in distorting the user's role when building the solution. In design literature, drawing a persona would be rather related to the Define phase \citep{mueller2018handbook}, but we noticed an anticipated use of this method to explore user needs at the beginning of the hackathon. We understood this as a way to compensate for the absence or limited access to stakeholders, and create hypotheses of the target users and their needs. Eventually, some participants would consider contacts within the event (participants and mentors) or outside (family, acquaintances, real stakeholders) to validate those hypotheses documented as a persona. Many interviewees already knew the method without any formal training on it, while others used it without that formal background. 
In the testimony from P7, we observed a group that never had any training to use that method and was not sure they were using it correctly: 
\begin{quote}
\textit{"sometimes, we didn't see things so well, then the mentor came and said: 'no, you think you don’t have a persona, but in fact, you do (have). This person here, you can draw a profile to solve the problem better'"}  - P7 \newline 
\end{quote}

\noindent{\textbf{Scenario}}. Also in an anticipated manner in the Discover phase to map user needs, we noticed some occurrences (P1, P20, P21, and P28) of what characterized the usage of scenarios -- although interviewees did not explicitly mentioned that term. In design, "a scenario is a concise description of a persona using a software-based product to achieve a goal" \citep{cooper2004inmates}. During interviews and direct observation of users, it is possible to learn about their tasks. It gives the opportunity to exercise empathy with users, which is a key concept for good design. Participant P20 reported on the usage of this method to build a narrative on usage scenarios of an imaginary system which was later validated in an interview over the phone. It helped them understand their target users would not utilize a mobile device of inappropriate size when using gloves:

\begin{quote}
\textit{"this allowed us to learn things about the professional activity and gave us direction to certain points of the project (...) We noticed the type of tablet to use in the application due to the use of gloves (for cleaning). These are types of details that we couldn't have known if we hadn't called them."} - P20 \newline 
\end{quote}


\noindent{\textbf{Empathy map}}. Two participants (P23 and P27) mentioned empathy maps as a way to identify users pains and gains, which can also be used to describe a persona \citep{ferreira2015designing}. In the interviews they indirectly mention how they try to speculate what these users see, hear or want (part of the typical guideline for an empathy map), characterizing the usage of the technique, although stating that they never had any formal training in design methods. Participant P27, for instance, explicitly mentioned the technique by its name and correctly summarized its concept, thus indicating he knows the method:
\begin{quote}
\textit{“in social hackathons, we use the empathy map a lot, to find out what they (users) see, how they feel, what they think, what they do... and give them a solution" - P27} \newline 
\end{quote}


\subsubsection{Define}

During this phase, there are practices that support participants finding the problem they will address. The main output of this stage is the problem statement, which will guide all the solution. The previous phase generates many inputs that have to be filtered and converged to a bold problem statement, which is one of the key aspects mentioned by winners of large and important hackathons. A well-defined problem will help to convince the jury in the end. Therefore, adequately filtering and choosing among the inputs of the Discover phase is paramount.

In Design Thinking, three constraints are frequently used to define an overlapping criteria for choosing successful ideas: \textit{feasibility}; \textit{desirability} and \textit{viability} \citep{brown}. They were all naturally mentioned in the interviewees' argumentation. Feasibility and desirability were the standard choices for them. They correspond to what is technically possible with the available resources (e.g., time, human resources) and what makes sense to users, respectively, whereas viability is a business-oriented criterion related to the commercial viability of the product. Although not always mentioning those constraints explicitly by name, the interviewees reasoning in the hackathons was usually guided by such criteria, showing evidence of a Design Thinking mindset. In the Define phase, the focus in on desirability, since participants are still framing the problem. Feasibility and viability would be considered later in phases where the candidate solutions emerge.  \newline

\noindent{\textbf{Affinity diagramming and card sorting.}}
The first phase (Discover) generates many inputs from the different methods that may have been applied. For instance, brainstorming or brainwriting generates many notes -- often physically detached as sticky notes. Therefore, it is important to apply other methods that help to converge those inputs from the previous phase. So, some participants (P5, P13,P15,P17,P18,P21, and P31) spontaneously used a range of clustering methods, applying mechanisms that closely resemble to affinity diagramming and card sorting \citep{hanington2012universal}. The former is a rather inductive process, where work is done from the bottom up, initially clustering specific, small details into groups, which will generate broader themes. The latter is a participatory design technique that allows exploring how participants group items into categories and relate concepts to one another. Their purpose is similar, and we find traces in some of the interviews, partly illustrated in the quote from P17.

\begin{quote}
\textit{"in the brainstorming, we wrote on post-it notes stuck on the wall. We removed the notes that were bad and put the good ones in another corner. In the end, we had several categories of post-its for technical problems, post-its about different themes"} - P17  \newline 

\end{quote}

\noindent{\textbf{Mind map and concept mapping.}}
In case the results of brainstorming are written on a whiteboard, paper or flip chart, the common approach is creating associograms using mind maps, which we identified being applied by participants P10, P16, P19, and P29. More elaborate mind maps resemble concept mapping method, which is in part illustrated in the following quotes from P19 and P21, respectively. Mind mapping helps to visually organize a problem space to better understand it and find creative and spontaneous association between ideas \citep{hanington2012universal, davies2011concept}. In its turn, concept mapping, which was used by P9, P14, P20, P21, and P25,  is slightly different and helps to outline relationships between ideas \citep{davies2011concept}, which we identified in the interviews as flow charts organizing the concepts. In the quotes from P19 it is worth mentioning a recurring democratic practice of common agreement based on majority voting, highlighted by many interviewees.

\begin{quote}
\textit{"a mind map on the whiteboard to connect what different people had provided in the brainstorming, collect the ideas on that board and converge them. For the selection (step), we tried to get everyone's opinion"} - P19
\end{quote}

\begin{quote}
\textit{"we made some stick notes over there, we took some markers and started writing over the table, over the board around us (...) So, the things we were facing, the issues, like multiple problems. We grouped the ideas. How can we do that, how can we do this. We did some mind maps and also developed some flowcharts"} - P21   \newline 
\end{quote}

It is worth noting that many participants mentioned a sort of "hack" in this process of choosing or converging ideas. They suggested (i) to use an explicit bias towards the evaluation criteria defined in hackathon rules as well as (ii) to try to guess the problems or domains that would please the members of the jury, as illustrated in the quote below. 

\begin{quote}
\textit{"if you want to win, you want to meet the hackathon evaluation criteria and please the judges. If the judge is a person who came from an environmental area, they will probably want to see something from that area in the application, because it will touch them directly. If the judge is from a company, they will want to see a clear business model"} - P6   \newline 
\end{quote}

\subsubsection{Develop}

This phase focuses on divergent thinking to generate ideas and filter them. There are many activities around prototyping here. Overall, many user experience and user interface concerns can be validated at this phase, which acts as an ideation phase that will result in the core concepts of the product to be developed. Having feedback on ideas and prototypes at this stage is fundamental, since any validation of utility or usability aspects with potential users is important to iterate on the prototypes and improve them. It allows early "pivoting", which happens when the team realizes a hypothesis is wrong, then updates the project and retests it \citep{muller2012design}. An interesting example of this sort of validation interviewee mentioned calling a hospital do discuss with doctors and nurses about features that made them see there were certain complications in the process that made them pivoting their idea. 

Although being a divergent phase, when prototyping and developing, some criteria for filtering and selecting ideas need to be taken into account. The evaluation criteria defined in the hackathon rules can be used to filter out their ideas. For instance, if the hackathon demands or gives extra points for using a certain API or technology, the solution would have to include that API, which to some participants seems to be forced. Feasibility, from the desirability-feasibility-viability triad \citep{brown}  is typically taken into account for choosing ideas, as mentioned by P5:

\begin{quote}
\textit{"Do what is achievable during that timeframe with the resources you have like available time, available software"} - P5
\end{quote}

When the hackathon is related to startups or business opportunities, the viability of the product is also included as a criterion. Because it is not necessarily a creativity technique, the Business Model Canvas \citep{osterwalder2010business} was not reported as one of the methods from Figure~\ref{fig:diamond}, but it appeared in many interviews (P5, P7, P13, P15, P16, P26, P27) as a tool to help to consolidate important concepts, such as target customers and value proposition.   \newline 

\noindent{\textbf{Brainstorming.}
Coming back to the design methods, brainstorming may take place again, supported by other methods that provide input for generating ideas targeting the solution or prototype to be developed, as described in many interviews (P14, P21, P23, P24, P26, P29, and P32) and illustrated below by P29:

\begin{quote}
\textit{"in terms of coming up with the ideas, we come up with buzzwords, like AI; we kind of juggle around with that word and connect the dots. This feature can do this, it can do that... By the end of that brainstorming session, we think of the different feedbacks and inputs from everybody in terms of workload or feasibility; in the end, we have a well-defined product that we can actually code up" } - P29   \newline 
\end{quote}


\noindent{\textbf{Competitive testing and SCAMPER.}}
A natural strategy unanimously adopted is to perform competitive testing or benchmarking \citep{hanington2012universal},  as evidenced in Table~\ref{table:design_methods} by a significant number of participants. In hackathons, the effort is concentrated on Web searches (e.g., Google), with many mentions to websites such as Product Hunt\footnote{www.producthunt.com}, to look for existing solutions that tackle the same target problem and quickly evaluate them. In this activity, one can analyze the strengths and weaknesses of products that are similar to what the group intends to develop. 

By identifying the limitations or shortcomings of existing solutions, the next step they naturally bring into a design process is the use of the SCAMPER, which was a practice confirmed in 16 interviews. It is a design method for idea generation \citep{mueller2018handbook} that can turn an existing idea into something new and different \citep{serrat2017scamper}. The SCAMPER method has been around for a while, is summarized as a mnemonic that consists on a flexible checklist of idea-spurring questions \citep{michalko2006thinkertoys}: Substitute something; Combine it with something else; Adapt something to it; Modify or Magnify it; Put it to some other use; Eliminate something; Reverse or Rearrange it. 

We observed that the usage of SCAMPER in hackathons helps to define or improving project features based on the analysis of competition. Even without knowing the existence of this method, the studied participants were typically guided by a subset of the verbs comprising the SCAMPER acronym. The quote from P15 summarizes the usage and combination of both techniques to generate project ideas:

\begin{quote}
\textit{"I usually like to Google that idea. There are some sites like Product Hunt, which are very good for finding what other people have done, and then search if there is a startup already doing something like that. There is a startup for everything (...) Most of the creative process is copying what other people have done and putting it on a different context; then, we add something on top of it or remove something. It is very difficult to come up with an original idea in a hackathon"} - P15    \newline 
\end{quote}

\noindent{\textbf{Minimum Viable Product (MVP).}}
The winners had a very clear understanding that they need to focus on an MVP, which the minimal set of features for solving the core
problem \citep{muller2012design}. The MVP is what is going to be presented to the jury. The concept is known and used by some interviewees (P5, P4, P7, P11, P20, P24, and P30), but in majority they though it as a common hackathon practice and did not now this concept was borrowed from the Lean Startup approach~\citep{ries2011lean} when asked. Participant P4 showed a clear understanding of the concept, as illustrated below:

\begin{quote}
\textit{"Once we have some rough idea, we then plan what is the most Minimum Viable Product. What do we need to deliver to showcase what we want to build? Often times is mainly from a user experience perspective, a user interface perspective."} - P4
\end{quote}

There were other interviewees completely unaware of the MVP term but very acquainted with its concept and the notion of its importance, showing a clear understanding of it as illustrated  in the counter-example of P23, who learned that proposing too many features and lacking focus is a bad practice:

\begin{quote}
\textit{"the errors that we had in two hackathons was not knowing what we were presenting. We did not have the idea of what product we had made, because we had too many functionalities and we had put many things when we were programming."} - P23    \newline 
\end{quote}

\noindent{\textbf{Low-fidelity and high-fidelity prototyping}}
Before starting the development of the product (i.e., a functional prototype), a good practice we identified in practically all interviews (Table \ref{table:design_methods}) was doing some form of prototyping \citep{hanington2012universal}, either low-fidelity or high-fidelity prototyping, which are very helpful to convey an idea and to quickly get user validation.
Most of the interviewees said they do at least some sketches on paper, as in the case of 24h hackathons where many times groups have little time to prototype before development. Typically, participants do some lo-fi prototyping with sketches on paper or whiteboards. Paper prototyping \citep{snyder2003paper} helps designing, creating and testing user interfaces. These prototypes are useful for discussing and quickly validating project ideas with mentors or other participants before jumping into development. In longer hackathons (e.g, 36h or 48h), groups may sometimes do lo-fidelity prototypes already using specific tools for prototyping mock-ups (e.g., Balsamiq). Some create hi-fi prototypes with tools specific for that (e.g., Figma, Sketch), simulating some interactions or providing look-and-feel details resembling the software to be developed. We observed that depending on the available time and performance of the team some groups would decide if both hi-fi and lo-fi prototypes or just on type would be done. The quotes of participant P9 illustrate that there is no rule of thumb:

\begin{quote}
\textit{"We made prototypes on the whiteboard a lot, talking to the UI designer to see what we could do. The final prototype was on Sketch (the tool), but it is also not a rule. Some screens we discussed on the whiteboard. In Sketch we thought of the final design."} - P9   \newline 
\end{quote}

\noindent{\textbf{Storyboard.}}
As another prototyping approach confirmed in some interviews (P1, P2, P3, P9, P11, P13, and P29), groups may also combine different wireframes or draw storyboards. 
A storyboard, which can be hand sketched in paper or on a whiteboard, helps understanding the flow of a user's interaction and how the interface will support each step \citep{snyder2003paper}. Storyboards were highlighted by P2 as being important with less experienced team members: 
\begin{quote}
\textit{"You show products based on user profiles, with storyboards that I do ... If there are junior people (in the team), a last minute change can technically be very expensive. So, we put it down on the paper in order to know how it is organized"}  - P2   \newline 
\end{quote}

\noindent{\textbf{Wizard of Oz.}}
In some hackathons, demonstrating a concept can be more important than completing its actual implementation. Participants may deliberately choose to fake some features -- although being explicit about not having implemented it -- just to illustrate them. Going in this direction, we found some interviewees (P18, P19, and P22) reporting on adapted usages of the Wizard of Oz method \citep{hanington2012universal, mueller2018handbook}, which consists of simulating system responses from behind the scenes. For instance, one interviewee stated that while some action was confirmed in the system, another group member sent a text message to the phone being used in the demo and a fake confirmation message popped up, simulating a push notification that was not implemented. Simulation is also useful in some Internet of Things projects, where sometimes the smaller scale models make necessary to fake some interactions because hardware may not be available, as in the example from P19:

\begin{quote}
\textit{"it is not because there are things that we are not able to do that we will deviate from this aspect or find an alternative solution. Is there anything that could replace that? We needed sensors, but we didn't have them. We simply simulated the operation as if there were sensors, without having them. Much of the proof of concept”} - P19   \newline 
\end{quote}

\subsubsection{Deliver}

The last convergence consists on delivering the functional prototype. The original Double Diamond suggests activities that make sense only if you are actually delivering the project to customers or users. As a Test Phase \citep{mueller2018handbook} where users would evaluate and give feedback.    \newline 

\noindent{\textbf{User testing.}}
Doing User testing \citep{da2011user} in the context of hackathons would actually be limited in terms of access to the actual users, unless domain specialists or hackathon sponsors who are stakeholders be present. Although the audience or potential users are sometimes contacted in earlier stages (discover and define phases), usually the case with user testing in this phase is doing simple tests with people who are in the same venue (mentors and other groups) to get feedback and, if possible, perform minor adjustments after that. User testing was a practice confirmed in more than half of the interviews, as displayed in \ref{table:design_methods}.

\begin{quote}
\textit{"We found the feedback from participants to be very useful. There's a lot more participants than mentors, so it’s easier to talk to them."} - P3   \newline 
\end{quote}

\noindent{\textbf{Pitch and storytelling.}}
The final moment is the pitch \citep{pincus2007perfect}, which consists in presenting the project to a jury. It was explicitly mentioned by almost all interviews (Table \ref{table:design_methods}). A few hackathons described by interviewees had a pre-pitch moment, which is a sort of presentation rehearsal for quick feedback before the final pitch. While some groups would just focus on delivering a pitch focused on features, a successful pattern for this is using a storytelling approach. Even if personas or scenarios have not explicitly been drawn in the first phase, they will likely appear here. The problem statement that resulted from the second phase was also mentioned as something important to be stated at this point, to attract the jury's attention. Similar to the choices that can be made in the previous phases to please the jury, we found the same pattern here where some participants present a
\textit{"tailor-made presentation based on the judges profile"} (P4).

As stated by P30, \textit{"The utility of the product is the most important thing"}. Although not using any more elaborate presentation technique such as using storytelling or a persona to support the pitch, he mentioned the successful usage of what he called attention getter, which would be something like impacting statistics or the impact of the target problem in society. Other interviewees mentioned that the pitch has to demonstrate that the prototype or product solves an important problem. One thing many participants mentioned is that storytelling is one of the approaches that make a difference in pitches, as in the example from P15:
\begin{quote}
\textit{"If there is something common in hackathons I’ve been is that good stories make great pitches. They really want to hear something that they can relate to.  So you have to tell them, 'this is me, this is what I think about that concept, once I had this kind of experience which was really bad, so I came up with this, in the hackathon, we did that and now we have this solutions' that’s the kind of basic thing I like to do.
"} - P15
\end{quote}

Storytelling was one common approach of the presentation of winning projects and was complementary to the pitch, as evidenced in interviews from P3, P15, P16, P17, P19, P21, P22, and P31. It is an approach familiar to designers, used in the conceptual design process to present concepts to clients \citep{parkinson2016engaging}. As Brown mentioned, \textit{"from the perspective of the design thinker, a new idea will have to tell a meaningful story in a compelling way if it is to make itself heard"} \citep{brown}. The statement made by P21 shows a lesson learned where he learned in practice a common design approach:

\begin{quote}
\textit{"If you think about the storytelling before starting to develop the product you might develop it much better. Because at one point at the time the story was going in a direction and the solution was not fitting into that. So you have planned the story before you can just align the product with that. "}  - P21
\end{quote}

\subsection{Enhancing the Process with Additional Design Methods}

During the focus group, participants proposed a set of additional design methods to enhance the design process identified in the interviews. In the Discover phase, participant P7 said groups can adopt \textbf{jobs to be done} \citep{christensen2016know, mueller2018handbook} to explore the origin of users’ motivations. In most cases, this method will support speculation, as the real user is often not in place during a hackathon. However, if it is possible to experience user situation and see the problem in practice, P5 mentioned hackathon groups can 
employ more \textbf{observations} \citep{hanington2012universal} to perform an ad-hoc ethnographic study (a lightweight version of design ethnography approach). 
Alternatively, they can benefit from 
the \textbf{5W1H} method \citep{andler2016tools, mueller2018handbook}, as pointed out by P3. Participants may use its structure (i.e., who, what, where, when, why, how) to think about the problem, in light of the theme or technology underlying the hackathon. Then, the groups can identify the real causes of a problem. 

The Develop phase can be enhanced by using \textbf{mood boards} \citep{hanington2012universal, mueller2018handbook} to organize the inspiration around the project and keep the ideas consistent with users goals and expectations, as mentioned by P8. He called this method \textit{``a living reference of the solution''}. The \textbf{User journey} \citep{hanington2012universal, mueller2018handbook} method was suggested by P7 as a way to check whether the proposed solution indeed brings value to the user routine. Additionally, the \textbf{morphological box} \citep{zwicky1967morphological}, cited by P2 and P4, would enable the groups to describe the problem in the absence of a wide dataset (real data is commonly scarce in hackathons) by using different dimensions. For instance, groups can consider ``motivation'', ``relevance'' and ``complexity'' as parameters.

\section{Discussion}

Considering our research question \textit{\rqone}, this section presents a discussion on expected and unexpected findings, based on our previous experience in the organization of hackathons.  Then, based on our data collection, we describe a set of recommendations from a design perspective based on the common practices adopted by hackathon winners. We conclude the section with a broader critical analysis of our findings.

\subsection{Expected and Unexpected Findings}
{Based on our own experience in the organization of hackathons, we expected some techniques and characteristics of the design process to be confirmed. Indeed, after the analysis of the interviews, there were some confirmations, but we were also surprised by some other practices that emerged for each phase, bringing aspects that were not foreseen. We discuss here the divergence and convergence principles that were noticed, followed by techniques that appeared as expected or not.}

Design Thinking brings the notion of divergent and convergent thinking. The divergence happens first, to generate and explore possibilities, followed by a converging phase to narrow down those possibilities to something more focused. As highlighted in Section \ref{DT_process}, a general model shows this structure happening twice, as a \textbf{Double Diamond} structure: first, diverging and converging on the problem space to understand the problem to be tackled; then, divergence and convergence are performed again in a phase focused on the solution ideation and prototyping. From our experience in hackathons, we see many groups jumping straight into the solution space (i.e., the second diamond), without much discussion on the problem to be solved, therefore skipping the first diamond. Groups that take the lead in these competitions usually (i) spend a good part of their time exploring the problem space, trying to understand the target audience of the challenge (hackathons usually have a theme) and their problems; (ii) converge to a problem statement, and only after that start (iii) speculate about solutions and possibilities to tackle that problem; to later (iv) converge to a final proof of concept that will be presented to the jury. We could confirm such double diverge-converge pattern in the interviews, where this strategy to approach a problem and think of a solution was clear. Some organizers and participants argued it is common to jump straight to the solution space in shorter hackathons (e.g., 12h). For instance, they can limit the problem to the event challenge (e.g., improve the quality of life of the visually impaired) without structuring any discussion or investigation about that problem and start the event by brainstorming an application (i.e., a solution) and its features. However, most of the interviewed participants highlighted the importance of even in shorter events spending some time discussing the problem, as it is fundamental to avoid solutions to problems without an audience or that does not exist.

During the \textbf{Discover} phase, divergent thinking takes place to explore the problem space and understand the audience, their needs and gather information about the problems. However, as we have observed in app development contests, participants usually build solutions based on their own perceptions without being based on user needs. This is a bad assumption that frequently takes place in these events and has been reported in literature \citep{lee2015open}. Based on our experience, we observed that some superficial exploratory Web research about the problem domain is typically done by savvy hackathon participants, to get acquainted with the domain. For instance, they may gather some statistics about the target problem (e.g., the percentage of the population who is overweight).  Nevertheless, we were amazed by the level of detail that some interviewees reported on their exploratory research. To our surprise, they mentioned the usage of quick questionnaires and interviews as resources to contact and gather data from their target audience. This was unexpected because hackathons tend to be isolated from the target audience or users but some groups reached out to people outside the event. We foresaw brainstorming to be mentioned by practically all participants since it is a widespread technique used very frequently in hackathons. Similarly, we also expected some brainwriting usage, but not as much as it was reported. One interesting aspect is that discussions to reach a consensus on the generated ideas was naturally merged as part of the brainstorming and brainwriting activities. We believe this is a probable reason for not having as many design methods in the Define phase as in the other phases, since the convergence was made together with a method used in the discovery phase. The usage of personas was not a surprise, as we have already seen occurrences in events we organized. However, we saw much more mentions to what characterized the usage of scenarios as a technique in this phase than we have observed from our experience as organizers. As a final technique we captured from interviewees in this phase, we were impressed by how naturally empathy maps were used to help the construction and description of personas.

In the \textbf{Define} phase brings a convergence of what was gathered in the discover phase, resulting in a problem definition. The general approach for the selection of alternative that we observed in our previous experience is around voting on topics, regardless of where they were written. We have already seen the usage of affinity diagramming and card sorting in events where we made sticky notes available. In the context of this research, some of the interviewees declared a more structured categorization approach when compared to what we had previously seen. Mind maps were not frequently observed in our context and we did not expect so much usage of this technique as it was reported in the interviews. The way on how concept mapping was performed in combination with mind maps was more elaborate than in the few times we had previously seen it. The major surprise in this phase was the interviewees allegedly observing the professional profile or background (e.g., sustainability) of the judges to use it as a criterion to influence the group choices aiming to take the project toward a direction that would please this audience (i.e., the jury).

In the \textbf{Develop} phase, divergent thinking takes place again. This time to generate options for the development of the project. The usage of brainstorming in this phase is very common, which we confirmed in the interviews. In the events we have organized, during the opening ceremony we use the term MVP (Minimum Viable Product) to explain that the goal in a hackathon is not to make a product full of features but with the feature or features that really matter for that problem. So, in our context, this was always common ground. The term was also known to many of the interviewees. Surprisingly, some participants of our interviews never heard of the term before but precisely described its concept, highlighting the importance of adhering to those principles in a hackathon. An important finding with interviewees concerned the natural way of combining benchmarking and SCAMPER. The level of detail on the recurring mention on searches on Google and Product Hunt in the pursue of apps to tackle similar problems was impressive in some interviews. By finding the potential competitors, participants instinctively adopted a SCAMPER-like approach -- without even being aware of that as a technique -- to help to propose something different from what exists in the market. The typical approach we were used to in our experience consisted of participants choosing a popular and widespread app as their benchmark and proposing their solution with a feature that app did not have. From our background, we witnessed groups frequently doing lo-fi prototyping in the form of sketches on paper and occasionally using specific software tools for lo-fi or hi-fi prototyping. This was confirmed in the interviews, but with the major difference on the massive usage of software tools for prototyping, especially in longer hackathons where they mentioned presenting well-crafted UI that was discussed and improved through prototypes. An interesting technique that emerged which we did not expect was that of storyboards. Its usage was reported in interviews as it typically is used in interaction design, helping to visualize a user’s journey or how the user would interact with the main features of an app. Some interviews declared the simulation of features through another person doing some action (e.g., an app confirmation message simulated as an SMS manually sent), consisting of a clear occurrence of the Wizard of Oz technique. This was unexpected for us since we have never seen its occurrence in any of the hackathons were involved.

\textbf{Deliver} would be the last phase on the double diamond, focusing on convergent thinking toward the delivery of the MVP to its audience. Some steps on validation. Since it is difficult to contact external users and there is a rush to finish the proof of concept, we assume some testing in this phase but rather among participants, as it was confirmed in the interviews. Although the pitch is a presentation technique expected in this sort of event, we noticed that some participants gave a high level of importance to it. There were many more mentions to presentation strategies than we expected, such as the straightforward way to a bold problem statement; the quick timing to get the jury's attention; the usage of impacting numbers; among other strategies. We reckon that storytelling was considered as an obvious choice for the pitch format but some interviewees did not appreciate that format and preferred what some of them called a more natural approach stating the problem and showing the intended benefits to users. This was unexpected for us since storytelling seemed to be a relatively frequent approach for pitching solutions in the hackathons we were involved. As hackathon organizers, what we also did not expect was the recurring "hack" of winners tailoring their solution or its pitch to please part of the jury, aiming at a better evaluation. \newline

\subsection{Recommendations to a Successful Participation in Hackathons}

An initial recommendation is that \textbf{teams must not start the hackathon with predefined ideas or solutions}. A na\"ive approach we observed in many groups is trying to start the project with a proposal in mind. Design Thinking is a user-centered approach that requires seeing the world through the user’s perspective. Hence, hackathon teams must employ methods to enhance their empathy with the audience (i.e., potential users), which is a core concept in design \citep{mueller2018handbook}. During \textit{Discover} phase, participants can explore divergent thinking and take a deep dive into the problem, since a weak exploration of the problem space may lead to solving an irrelevant problem. The importance of a clear and relevant problem statement was repeatedly mentioned among winners, as it facilitates the transition to a potential solution. Such immersion phase is strongly related to requirements engineering, as it aims to get a minimum understanding of the domain and users \citep{mendez2019}.

A second recommendation is that \textbf{groups must plan for the timeframe, rules and hackathon jury}. The dedication to the problem and solution spaces will vary according to the timeframe of the event, usually 24h to 48h but may also vary from 24 hours to five working days \citep{valenca2019}. For instance, a one-day event requires quicker decisions with more focus on the solution space, while a two or three-days hackathon allows more exploration in the problem space. Once a solution is developed, it is assessed by judges via team pitches. During the focus group, participants highlighted the relevance of understanding the evaluation process. The criteria may be described in the event regulation or simply resides in the judges minds, requiring an analysis of their profiles. This strategy represents a design heuristic: aligning the project's desirability with the criteria that will be used by organizers.

A third recommendation concerns \textbf{identifying existing solutions to propose more innovative ones}.
The team should not repeat existing products, but rather improve them in certain aspects. A combination of benchmarking (i.e., competitive testing) with the SCAMPER~\citep{serrat2017scamper} technique can help in that direction. The first step is identifying what exists through quick Web searches or using specific engines focused on digital products (Product hunt, App Store, Google Play, etc) in order to identify solutions that already tackle the target problem. Then, the usage of SCAMPER can help to suggest some addition to or modification of the identified application or solution that already exists. 

A fourth recommendation concerns \textbf{prototyping to validate ideas before the team starts constructing the solution}. Prototypes help to convey ideas and are useful to have quick validation from team members and potential stakeholders \citep{klotins2019software}. Even paper prototypes can be very helpful to design and test user interfaces \citep{snyder2003paper}. In the case of the impossibility of contacting such stakeholders, people participating in the hackathon, especially mentors (technical or domain mentors), can be very useful to help to validate prototypes. 

As a fifth recommendation, \textbf{the pitch has to attract the jury's attention}. In hackathons, as well as in professional life, communication skills are very important. Conveying an idea through a story and attracting the attention of your audience is key. In addition to storytelling and rhetoric, even body language is important in order to do a convincing pitch \citep{torres2019you}.  There are basic guidelines to be followed, such as the pitch being short, start with a winning logline and emphasize on one thing the presenter wants their audience to remember in that pitch \citep{gallo2018pitch}. Regardless of a storytelling format being used or not, the team needs a bold problem statement and engage with the target audience to show the relevance of the product that was developed and is being presented. A good project without a good pitch would be in disadvantage to an average project with a good pitch.

Finally, we suggest that \textbf{organizers should raise participants' awareness about design thinking}. Although organizers typically do not impose any design thinking process, when they rarely suggest any format it follows a design thinking philosophy. As this approach is reinforced, with proper training sessions (e.g., tutorials previous to the event), the process is understood by participants and becomes more repeatable or reusable. By employing design thinking as a complementary process to support software development, participants explore their problem-solving ability to make their outcomes more innovative. However, any process or design method should be a support tool instead of a strategy imposed by organizers. This is important to avoid the feeling of losing the freedom that some participants may face when forced to adhere to imposed practices.

\subsection{Critical Analysis}

Our investigation revealed that hackathon participants hold a clear notion of the importance of a user-centered approach and use many design methods without being aware of that. However, we found a deviation of this user-centered approach when participants put the jury members in the central perspective as a user. It was a strategy to please the jury in an attempt to get more attention and better scores in their project evaluation.

We also observed an alignment of culture and practices to what is seen in software startup companies that can also bring advantages to the software engineering field as a whole.  \cite{lindberg2011design} supports the idea of applying design thinking to IT development (i.e., hardware and software development) as a complementary thinking style that can help IT development teams to produce more innovative results. Additionally, women can enhance the design skills and innovative capacity of a team, as disruption in research and development tend to come from more gender diverse groups \citep{diaz2013gender}. This innovative problem-solving approach could be leveraged  by broadening the participation of female members in hackathons. But these events tend to have a low number of women participating due to its competitive and predominantly sexist environment  \citep{paganini2020engaging}. This reality of women's underrepresentation was reflected in our study, as the participant winners of hackathons we interviewed were overwhelmingly men (Table 3). 

In general, the awareness of design methods and processes practiced in hackathons can generate many benefits for IT developers:

\begin{itemize}
 \item \textbf{The startup entrepreneurial process underlying a hackathon}: we identified that the conception and development of software in hackathons resemble, on a smaller scale, the software engineering approach that takes place in software startups. Similar to what was described by interviewees, in startups the value proposition (target audience, promised benefits and competitive features) of the MVP is used to identify software requirements is one of the first steps in product engineering activities \citep{klotins2019software}. The culture of pitching \citep{gallo2018pitch,pincus2007perfect} ideas and products is also a key characteristic in the startup world that is also present in hackathons. In fact, the overall approach taken by software startups is very much related to what is advocated by design thinking and lean startup methodologies, which are not only processes but also include tacit knowledge in the form of practices and mindsets \citep{muller2012design}. Therefore, hackathons foster an entrepreneurial spirit towards new business opportunities. 
 
 \item \textbf{The close relationship between the hackathons and agile requirements engineering}: the design methods identified in hackathons resemble the approach taken in the software industry to understand requirements, especially in the context of software startups where agile practices take place. Practitioners use interviews, surveys, observations and demonstration of prototypes to discover new requirements, and validate existing ones \citep{klotins2019software}. On their turn, hackathon competitors consider people from their teams and their social network as requirements sources, while product value proposition determines what should be prioritized. In addition, such reports on startups suggest these companies typically using brainstorming, mock-ups and wireframe to understand requirements and design user interfaces of products, aiming at quicker requirements validation, in a similar way to what teams do in hackathons.

 \item \textbf{The possibility to foster a new mindset via hackathons}: software development companies face challenges in integrating design methods and processes into software methodologies~\citep{ogunyemi2019systematic}. By adopting and incorporating Design Thinking methods into their software development culture, these hackathon participants take a step forward in regards to other professionals from the field. These participants can shift from the traditional software engineering mindset to cultivate a new creative mindset that is critical for software companies \citep{muller2012design}. Hackathons are shaping a new culture of IT engineers that are spontaneously incorporating a design thinking mindset. Hence, hackathons act as eye-opening events that can help companies to establish a result-driven culture that promotes innovative, solution-oriented and self-directed thinking in developers \citep{valenca2019}.
\end{itemize}

\section{Final Considerations}
\subsection {Contribution}

{We believe our work helps to complement existing literature about hackathons, which already explored some design thinking perspectives on such events but lacked details about the ideation practices used in hackathons. Considering the context of competitive focus-centric hackathons and based on common practices followed by worldwide participants who are frequent winners, we identified and enumerated the recurring design methods and strategies that seem to lead to successful hackathon participation. We positioned those methods under the phases of the Double Diamond design process model. By bringing a Design Thinking perspective, we presented how these design methods were applied following divergent and convergent thinking in different moments of this process. We also provided recommendations for successful participation in hackathons as well as additional design and creativity methods that can be put into practice. In addition, we draw a parallel between the dynamics of the design thinking approach in hackathons with the software engineering practices conducted in software startups. Our analysis revealed a shared mindset of professionals who understand the importance of design, which is not based on random inspiration moments but rather in the usage of methods and hard work}. 

By shedding light on the methods adopted throughout this structure, we provide relevant inputs to approach the topic in future studies. We enable other researchers to apply our process perspective and verify its results in a given scenario, via an improvement of the case study or through other methods such as action research. 
\subsection {Limitations and Threats to Validity}

A possible threat to \textbf{internal validity} lies in our restriction to the perspective of hackathon winners. {Our experience in the field made us notice the strong usage of design methods as a recurring pattern in winning groups, however only selecting winners for interviews generated a bias in the sample. Therefore, we cannot claim that such a design thinking approach is only followed by winners. We did not collect data to verify that other groups not necessarily having a winning history could share such design practices to the same or to a lesser extent than winning groups; or if perhaps there are typical design methods (e.g., brainstorming) normally used by all groups, but not necessarily situated in an informal Design Thinking process. By interviewing the other set of participants of these events, we could speculate about possible uses of design methods and their impact of the results presented by the teams. Such correlation, with a cause-consequence clause (e.g., a given design process generates a given result), is rather associated with a quantitative analysis of the phenomenon. Another limitation of our work is that we investigated only the context of competitive focus-centric hackathons. Other settings, such as non-competitive events or hackathons that are less innovation-oriented and more social-focused, which are mostly multidisciplinary, may lead to different sets of practices or design methods employed by participants.}

{We also acknowledge a possible threat to \textbf{construct validity}. Our research did not follow an exploratory stance when we would find out what was going on and seek new insights into a specific emergent phenomenon. Instead, we aimed at depicting a design process we perceived as being applicable to many hackathons. Hence, we may have translated some of our initial hypotheses into interview questions, but the rounds of discussions among authors during the construction of the interview guide helped to avoid leading questions or questions that were very specific to what we have experienced in hackathons. We tried to keep them neutral (e.g., participants were not supposed to choose among options) and rather coarse-grained. This was a means not to restrict or direct participants.}

The \textbf{conclusion validity} of our research may have been threatened by two main factors. First, the live coding procedure provided us with a non-exhaustive transcription, i.e. we may have neglected important data while hearing the audio records from interviews. To reduce such threat, we guaranteed that the researchers in charge of conducting the interviews were also responsible for listening to the audio files. Their familiarity with the data enabled us to guarantee a proper selection of important excerpts. {Second, part of the authors had a large experience in organizing hackathons and a strong background in Design Thinking. This shared knowledge on that topic among the authors had a risk of confirmation bias to support the initial hypothesis. Hence, we faced the risk of introducing such experience in the results and general insights, i.e. our interpretation of the phenomenon would also stem from non-systematic observations naturally performed during other hackathons. To reduce the chances of having different types of researcher bias (e.g., confirmation bias, leading questions) on top of the actual data, we promoted discussions among researchers during the analysis or right after deriving a set of conclusions. This approach allowed us to have a less biased understanding of interviewees discourse, with an analysis based on objective data, although we acknowledge that it may have not completely eliminated researcher bias since all authors shared a Design Thinking perspective that may have persisted despite the discussions. Moreover, it mitigated translation issues, since we had four different idioms in our dataset. We decided to present our results to a group of people experienced with hackathons via a focus group as a means to verify the process generated from collected evidence. This whole procedure strengthened our data analysis.  The fact of this focus group be composed of people from the professional network of the authors may have introduced personal bias with a tendency to agreement on the proposed structure of phases and respective techniques situated on each phase. Nevertheless, they were asked to criticize the model which resulted in suggesting additional techniques. } 

\subsection {Future Work}

We plan to conduct the following studies in future research:

\begin{itemize}
  \item \textbf{Performing an ethnographic study:} we plan to perform ethnographic studies to observe \textit{in loco} how hackathon participants interact with design methods. This method is essential to address the common lack of details in an interview, mainly because participants may forget or simply omit (considering a given information as irrelevant) practices that could enrich our analysis. Hence, ethnography will allow us to observe teams dynamics in a hackathon. In particular, it can raise our understanding of whether methods were used in a correct manner or with needed changes due to constraints of the hackathon. By suggesting the overarching double diamond design process model as a starting point to different and specific forms of hackathons (e.g., educational hackathons conducted by university professors, or civic hackathons held by public organizations), we can analyze the relationship between specific demographic factors (e.g., duration and location of the event, background of participants, and even gender of participants) -- not approached by this paper -- and the design methods used by hackathon participants. Such a sociological view of the phenomenon will contribute to studies on human factors in software engineering. 
  \item {\textbf{Extend the study introducing a gender perspective:} in the first round of data collection, our pool of interviewees was majorly formed by male participants (only one woman among the thirty-seven participants). The same for the focus group, when we had six men in a group of eight participants. The literature denotes that teams with a higher proportion of female members have better results on a collective intelligence factor than other teams. \citep{woolley2010evidence}. Hence, it is important to have different profiles properly represented in the dataset. A similar study that brings the gender dimension could explore the creativity factor that is involved on gender-diverse teams and what role it plays on the design methods that are used by those teams in comparison to male-only groups.}

\end{itemize}

\bibliography{submission}

\section*{About the authors}

\textbf{Kiev Gama} is an associate professor of
computer science at the Universidade Federal de Pernambuco (UFPE), Brazil. He received his PhD in computer science from the University of Grenoble,
France. His research interests include several aspects of time-bounded collaborative events (e.g., hackathons, game jams) within contexts such as open innovation and education.

\noindent\textbf{George Valença} is a professor of software engineering at Universidade Federal Rural de Pernambuco (UFRPE), Brazil. He received an MSc and a PhD in Computer Science from Universidade Federal de Pernambuco (UFPE), Brazil. He researches on open innovation and business process management approaching topics such as power relationships, data protection and corporate hackathons. 

\noindent\textbf{Pedro Alessio} is a professor at the Department of Graphic Expression at Universidade Federal de Pernambuco (UFPE), Brazil. He holds a bachelor's degree in Industrial Design from the Federal University of Pernambuco, a master's degree and a PhD in Informatics from the Conservatoire National des arts et métiers (CNAM), France.

\noindent\textbf{Rafael Formiga} holds Master's degree in Design from the Universidade Federal de Pernambuco (UFPE), Brazil. He is a product designer at VTEX. His main research focuses on design methods, product design and UX research. He works with digital design education, building bridges between academia and industry.

\noindent\textbf{André Neves} is an associate professor at the Universidade Federal de Pernambuco (UFPE), Brazil. He has a bachelor's degree in Industrial Design from the Universidade Federal da Paraíba (UFPB), and a Master's and PhD in Computer Science from UFPE. He is experienced with the Design of computer systems, working mainly in the investigation, development and application of methods and design techniques as an instrument of innovation.

\noindent\textbf{Nycolas Lacerda} graduated in computer science from the Universidade Federal Rural de Pernambuco (UFRPE), Brazil. His main research focus is on corporate hackathon organizations, having participated in a research group "Innovation for Business and Software Evolution - IBSE" as a student researcher. He currently works as an IT consultant.

\appendix

\section{Interview Guide}

\begin{enumerate}
    \item  \textbf{Demographic Information}
    \begin{itemize}
        \item Age
        \item Educational background
        \item Time of experience in the area
        \item Experience with hackathons (total participations, prizes won)
    \end{itemize}
    \item \textbf{Hackathon Dynamics}
    \begin{enumerate}
        \item Groups
        \begin{itemize}
            \item How groups were formed?
            \item Was there interdisciplinaraity?
            \item What was you role in the group?
        \end{itemize}
        \item Creative process
         \begin{itemize}
            \item Did the organizers / mentors suggest the use of any creative technique for idea generation?
            \item Did you know / use any specific Design Thinking methods?
            \item What was the step-by-step process that your team followed to design and develop applications in hackathons?
            \item Did you work with someone from design domain? How did he/she influence the project?
            \item How were the tasks managed?
            \item Did you already have any idea for the hackathon project? If yes, how was it generated?
            \item Could you identify or have access to the target audience or domain experts?
            \item Were there any guidelines to generate and select the application solution alternatives?
            \item Did you research about potential competitors of your solution?
            \item Did you think of a differentiation strategy in relation to competitors?
            \item Did you do any low fidelity (lo-fi) prototype?
            \item Could you do any validation of the prototype or application with users or experts?
            \item How tests were handled?
            \item How the pitch was made?
        \end{itemize}
    \end{enumerate}
\end{enumerate}

\end{document}